\documentclass[11pt]{article}

\usepackage{epsfig}
\usepackage{amsfonts,amssymb,amsmath}
\usepackage{vmargin}
\usepackage{cite}

\numberwithin{equation}{section}

\newcommand{\A}{\ensuremath{{\cal A}}}

\newcommand{\D}{\ensuremath{{\cal D}}}

\newcommand{\F}{\ensuremath{{\cal F}}}

\newcommand{\N}{\ensuremath{{\cal N}}}
\renewcommand{\O}{\ensuremath{{\cal O}}}

\newcommand{\Q}{\ensuremath{{\cal Q}}}

\newcommand{\tA}{\ensuremath{{\tilde{A}}}}
\newcommand{\tB}{\ensuremath{{\tilde{B}}}}
\newcommand{\ta}{\ensuremath{{\tilde{a}}}}
\newcommand{\tb}{\ensuremath{{\tilde{b}}}}

\newcommand{\rf}{\ensuremath{{\bar{f}}}}
\newcommand{\rh}{\ensuremath{{\bar{h}}}}

\newcommand{\ra}{\ensuremath{\rightarrow}}

\newcommand{\half}{\ensuremath{\frac{1}{2}}}

\newcommand{\quarter}{\ensuremath{\frac{1}{4}}}

\newcommand{\be}{\begin{equation}}
\newcommand{\ee}{\end{equation}}

\newcommand{\ba}{\begin{eqnarray}}
\newcommand{\ea}{\end{eqnarray}}

\newcommand{\ns}{\normalsize}

\newcommand{\gsim}{\raise.3ex\hbox{$>$\kern-.75em\lower1ex\hbox{$\sim$}}}
\newcommand{\lsim}{\raise.3ex\hbox{$<$\kern-.75em\lower1ex\hbox{$\sim$}}}

\newcommand{\nn}{\nonumber}

\newcommand{\w}{\wedge}

\newcommand{\Vol}{\mathcal{V}}

\newcommand{\im}{\mathrm{Im\;}}
\newcommand{\re}{\mathrm{Re\;}}

\newcommand{\href}[1]{\underline{#1}}

\newcommand{\vp}{\varphi}

\begin{document}
\begin{titlepage}
 
\title{
   %\hfill{\ns hep-th/xxxxxxx\\}
   \vskip 2cm
   {\Large\bf Low Energy Supersymmetry from Non-Geometry}
   \\[0.5cm]
   {\ns\Large  Eran Palti \footnote{email: palti@thphys.ox.ac.uk} }
   \\[0.5cm]
   {\it\ns Rudulf Peierls centre for Theoretical Physics, University of Oxford}\\
   {\it\ns Keble Road, Oxford, UK. } \\[0.2em] }
 
\date{}
 
\maketitle
 
\begin{abstract}\noindent
We study a class of flux compactifications that have all the moduli stabilised, a high (GUT) string scale and a low (TeV) gravitino mass that is 
generated dynamically. These non-geometric compactifications correspond to type II string theories on $SU(3)\times SU(3)$ structure orientifolds. 
The resulting superpotentials admit, excluding non-perturbative effects, supersymmetric Minkowski vacua with any number of moduli stabilised. We argue that non-perturbative 
effects are present and introduce terms in the superpotential that are exponentially suppressed by the same moduli that appear perturbatively. These deform the 
supersymmetric Minkowski vacua to supersymmetric AdS vacua with an exponentially small gravitino mass. The resulting vacua allow for low scale supersymmetry breaking 
which can be realised by a number of mechanisms. 
\end{abstract}
 
\thispagestyle{empty}
 
\end{titlepage}

%%%%%%%%%%%%%%%%%%%%%%%%%%%%%%%%%%%%%%%%%%%%%%%%%%%%%%%%%%%%%%%%%%%%%%%%%%%%%%%%%%%%%%%%%%%%%%%%%%%%%%%%%
\section{Introduction}
%%%%%%%%%%%%%%%%%%%%%%%%%%%%%%%%%%%%%%%%%%%%%%%%%%%%%%%%%%%%%%%%%%%%%%%%%%%%%%%%%%%%%%%%%%%%%%%%%%%%%%%%%%

Low energy (TeV) supersymmetry is an attractive solution to the hierarchy problem. Although string theory naturally incorporates supersymmetry 
at the string scale it is still not clear why it should only be broken at a much lower scale. The success in moduli stabilisation 
make this question one that can be studied concretely.  
A particularly attractive idea is that supersymmetry is broken at a dynamically low scale \cite{Witten:1981nf}.  In terms of 
moduli stabilisation a necessary condition for this is that the moduli are fixed at a point with a dynamically low gravitino mass.   
Consider the KKLT set up where the Standard model is realised on some brane which may be located
in a warped region \cite{Kachru:2003aw}. Then the gravitino mass and the string scale are schematically given by \cite{Burgess:2006mn} \footnote{These expressions are modified in the warped case 
with the volume scaling to a different power. However since in that case we consider the volume to be of order one this is not important for 
illustration purposes.}
\be
m_{\frac32} \sim \frac{M_p\;e^A\;W_0}{\Vol} \;,\;\; m_s \sim \frac{M_p\;e^A}{\Vol^{\frac12}} \label{gravschem} \;,
\ee
where $M_p$ is the Planck mass, $W_0$ is the integrated out complex structure moduli superpotential, $e^A$ is the warp factor at the brane, 
and $\Vol$ is the Calabi-Yau (CY) volume. There are realisations of scenarios where either the volume is exponentially large \cite{Balasubramanian:2005zx,Conlon:2005ki} or where the warping is exponentially small \cite{Giddings:2001yu} both leading to a dynamically low gravitino mass. However the scenarios also imply an intermediate and low string scale respectively.  If we wish to keep a high string scale then a simple solution is to take  
$W_0$ hierarchically small. This scenario already has a realisation in string theory by fine tuning the string scale numbers in $W_0$ to cancel to hierarchical accuracy. 
The aim of this paper is to explore an idea for generating a {\it dynamically} low $W_0$.  

The obvious way to get a dynamically low $W_0$ is to find vacua where $W$ vanishes up to non-perturbative effects that are exponentially suppressed.    
This requires two important ingredients. The first is that in the vacuum the superpotential vanishes (including all the perturbative corrections to it that arise away from 
the large complex-structure and large volume limit). The second is that 
the moduli appearing in the exponential non-perturbative terms are already perturbatively fixed at some, ideally parametrically controlled, large value. This is the mechanism that generates the hierarchy scale. 
A possible realisation of this type of scenario has already been explored in string theory. It works as follows. We consider type IIB string theory compactified on a CY which has a single complex-structure modulus. 
We also include R-R and NS-NS flux. It was shown in \cite{DeWolfe:2004ns} that it is possible to pick the fluxes such that the complex structure modulus is fixed at the Landau-Ginzburg (LG) point and the perturbative superpotential vanishes exactly. 
We further consider gaugino condensation on a D3 brane on a singularity in the CY which induces a non-perturbative term in the superpotential that is exponentially suppressed by the value of the dilaton. Since the dilaton is already fixed perturbatively by the fluxes this exponential can generate a hierarchy deforming the vacuum to one where $W_0$ is dynamically small. The Kahler moduli are then fixed through the usual KKLT \cite{Kachru:2003aw} mechanism. 

We now summarise some of the downfalls of this construction. The first one is that this stops working once the CY has more than one complex-structure modulus. Attempts to implement this scenario for the more 
generic cases of multiple complex-structure moduli \cite{DeWolfe:2005gy} have only been successful at the cost of leaving flat directions in the complex-structure moduli space. Since we are interested in constructions where all the moduli are 
stabilised this is unsatisfactory\footnote{It was suggested in \cite{DeWolfe:2005gy} that it may be possible to fix the remaining moduli through their appearance in multiplying non-perturbative effects in the Kahler moduli. It would be interesting 
to explore this possibility.}.  Another problem is that it is completely reliant on the D3 gaugino condensation.
A phenomenological problem is that since the Kahler moduli only appear non-perturbatively, they get fixed at values where their vacuum mass is of order the gravitino mass which implies a light moduli problem. Like the warping and large-volume models, these constructions are only valid within a type IIB scenario meaning that we still have no method to generate a small gravitino mass within geometric compactifications of IIA. Finally, it was suggested in \cite{Dine:2005gz} that these vacua are very rare in the landscape.                 
In this paper we argue that the introduction of non-geometric fluxes can potentially solve all of the aforementioned problems. Given the lack of explicit constructions 
of non-geometric compactifications, we can only study toy models that show how the issues may be resolved. We outline below the key changes that non-geometric fluxes imply 
for the geometric scenario previously outlined. 

Non-geometric fluxes are flux parameters that are associated with the failure of a manifold to be globally patched 
using only geometric symmetries. Rather they correspond to the manifold being globally patched using T-dualities. For some exploration into the geometric interpretation of non-geometric fluxes see \cite{Dabholkar:2002sy,Kachru:2002sk,Hellerman:2002ax,Flournoy:2004vn,Hull:2004in,Shelton:2005cf,Aldazabal:2006up,Ellwood:2006ya,Grana:2006hr,Micu:2007rd,Marchesano:2007vw,Ihl:2007ah}. 
Although such fluxes seem quite exotic, in many ways they should not be regarded as so any more than NS flux. 
The reason is that they are precisely the mirror of NS fluxes. So that NS flux in type IIB string theory leads by 
mirror symmetry to non-geometric fluxes in type IIA. Therefore what seems exotic in one corner of M-theory is quite natural in another corner. However, 
in this paper we consider theories that can not, by any dualities, be 
brought to a frame where they correspond exactly to a compactification on a CY with just (RR and) NS flux. Since we have no explicit examples we can only assume that the different fluxes can be superposed. There is no known reason forbidding their superposition and it may be that with better understanding of non-geometry we can develop explicit examples that realise the toy models studied in this paper\footnote{It is interesting to note that after the superposition of fluxes the theories admit vacua, found in \cite{Micu:2007rd}, that violate the conjecture in \cite{Becker:2007dn}. They are vacua where there is a flat direction in flux space where the mass of the moduli does not reach down to the AdS scale. If it could be shown that the conjecture of \cite{Becker:2007dn} holds for all solutions of string theory then this would rule these theories out. 
We thank K.Becker, M.Becker and J. Walcher for discussions on this point.}. The purpose of this paper is not to study whether the superposition of non-geometric fluxes is viable or not. Rather, given that there is some motivation for their existence, the aim is to study how their inclusion affects some of the results that apply to the geometric constructions studied so far in the literature. 

Non-geometric fluxes induce new terms in the superpotential. We discuss these in section \ref{nongeocy}, and in section \ref{susymink} we show that the new terms allow for constructions where all the moduli are fixed, with a string scale mass, in a, up to non-perturbative effects, supersymmetric Minkowski vacuum. 
This vacuum is exact in the sense of corrections away from the large complex-structure limit in IIB or their mirror $\alpha^{'}$ corrections in IIA. 
For the form of these corrections we consider the periods to take the form of CY periods. This is an assumption since it could be that the introduction of the non-geometric 
fluxes modifies these corrections to the superpotential. However we argue that mirror symmetry implies that such an assumptions is not completely unreasonable. For example 
the mirror to a IIB CY with H-flux is non-geometric but, by mirror symmetry, the corrections to the superpotential are of the form dictated by the CY periods. 
We present some arguments that these may be the only perturbative corrections that the superpotential receives and in that sense the vacua are perturbatively exact\footnote{In the case of supersymmetric Minkowski vacua corrections to the Kahler potential do not affect the vacuum.}. However it could 
be that there are further perturbative corrections to the superpotential that we do not consider in this paper. In that case it would require further study to 
determine if vacua that are also exact in terms of these corrections can be constructed. For the case of the corrections studied, 
 we discuss two different methods by which Minkowski vacua can be realised and give two toy examples based on CY data. We give some evidence that at least the complex-structure moduli in IIB or the Kahler moduli in IIA may get fixed at parametrically controlled values. We also argue that within the non-geometric landscape these type of vacua are not as rare as in the geometric landscape.

In section \ref{nonper} we study whether we expect non-perturbative effects to be present in non-geometric compactifications. We show that strong supergravity gauging constraints and world-volume anomalies allow for the presence of these effects. In particular they allow 
for exponential terms in the superpotential that are functions of the same superfields that appear perturbatively. This realises the mechanism of generating a hierarchy.
We also discuss how intersecting branes constructions may require axions from the closed string sector to cancel anomalies. This requires turning off some 
of the fluxes to destabilise a modulus. These kind of constructions are a type of generalised KKLT setup where the moduli are split into a sector that is fixed perturbatively 
and one that is fixed non-perturbatively. We argue that these have most of the general features of KKLT and in particular they suffer 
from light moduli in the case of a low gravitino mass. 

Finally in section \ref{susybreak} we discuss some models of supersymmetry breaking that still maintain a small 
gravitino mass. A particularly attractive mechanism for this is supersymmetry breaking in a metastable vacuum of an open string gauge sector. 
Non-geometric fluxes imply this sector can be realised on D3 or wrapped D6 or D7 branes. They can also fix all the moduli at a high string scale 
allowing for gauge mediation without a light moduli problem. We summarise our findings in section \ref{summ}.    

%%%%%%%%%%%%%%%%%%%%%%%%%%%%%%%%%%%%%%%%%%%%%%%%%%%%%%%%%%%%%%%%%%%%%%%%%%%%%%%%%%%%%%%%%%%%%%%%%%%%%%%%%
\section{Non-geometry and Calabi-Yaus}
\label{nongeocy}
%%%%%%%%%%%%%%%%%%%%%%%%%%%%%%%%%%%%%%%%%%%%%%%%%%%%%%%%%%%%%%%%%%%%%%%%%%%%%%%%%%%%%%%%%%%%%%%%%%%%%%%%%%

In this section we motivate an extension to the Gukov-Vafa-Witten (GVW) IIB superpotential that results from 
compactifications on CY manifolds with flux. The extension corresponds to turning on non-geometric fluxes. 
We are particularly interested in the form of the superpotential away from the large complex-structure large volume limit since 
we are hoping to find perturbatively exact Minkowski vacua. The presence and effects of non-perturbative corrections 
is studied in section \ref{nonper}. We also discuss the mirror IIA superpotential and the corrections that it 
receives. We begin with an introduction to non-geometric fluxes. 

The idea of non-geometry arises naturally in the context of T-duality with flux. It was shown in \cite{Kachru:2002sk} starting from a torus with some NS H-flux 
through it and performing a T-duality we reach a torus with twisted boundary condition, a twisted torus. The twisted torus has no H-flux through it, 
but instead is no longer flat but has some torsion. The torsion is parameterised in terms of metric fluxes which 
are the parameters that the original H-flux has transformed into. If we now perform another T-duality we reach a `torus' that is no longer periodic up 
to any geometric transformation. Rather we find that it is periodic up to T-dualities, it is a T-fold, or a non-geometric `manifold' \cite{Dabholkar:2002sy,Hellerman:2002ax,Flournoy:2004vn,Hull:2004in}. Since all we have 
done is performed some T-dualities, which is a symmetry of string theory, this background should be just as valid as a string background as the original torus with H-flux. 

Consider compactifying type IIB string theory on the original torus with H-flux to recover an effective four-dimensional theory. 
We may ask what happens to the effective four-dimensional theory as we perform the T-dualities? 
Since type IIB and type IIA string theories are related by a T-duality, we expect that compactifying type IIA string theory on a twisted torus should lead to the same four-dimensional theory. This was studied in a number of cases \cite{Kachru:2002sk,Derendinger:2004jn,Villadoro:2005cu,Camara:2005dc,Grana:2006kf} and shown to hold. 
It is therefore reasonable to conjecture that the same should hold for a 
compactifications on a non-geometric torus. This has lead to a number of studies of the possible effective potentials that could arise from such 
compactifications \cite{Shelton:2005cf,Aldazabal:2006up,Ihl:2007ah}. Moduli stabilisation was subsequently studied using these potentials where it was shown that generically all of the moduli are fixed. However we should note that in all the cases studied, once the tadpole equations were imposed, the vacua where all the moduli were fixed 
were always AdS. 

Now consider CY compactifications. It has long been realised that mirror symmetry relates the four-dimensional effective theories resulting from compactifications of type IIA and IIB string theory on two mirror CYs. 
The SYZ conjecture \cite{Strominger:1996it} is that this symmetry should be interpreted as performing three T-dualities in the internal CY. 
Mirror symmetry has been very precisely tested. It quantitatively relates the 
two mirror four-dimensional theories up to calculable perturbative and non-perturbative corrections \cite{Candelas:1990rm,Candelas:1993dm,Candelas:1994hw}. 
The interpretation of mirror symmetry as T-duality leads to the expectation that it should also hold for consistent string backgrounds with flux. 
Consider compactifying IIB on a CY with H-flux. Mirror symmetry is three T-dualities and we expect mirror theories corresponding to 
compactifications on four different types of geometries corresponding to having flux along zero, one, two or three of the T-duality directions. 
Depending on what type of H-flux we have present the geometry of the mirror manifold changes. For the case of zero we recover a mirror CY manifold. 
The case of one leg corresponds to having what is termed electric H-flux. Then the mirror theory results from a compactification of type IIA string theory 
on a manifold with torsion, the general class that preserve the $N=2$ supersymmetry of the CYs are termed manifolds with $SU(3)$ structure and 
compactifications on these general manifolds were studied in \cite{Gurrieri:2002wz,Gurrieri:2002iw,Tomasiello:2005bp,House:2005yc,Grana:2005ny,Micu:2006ey,Gurrieri:2004dt,Micu:2004tz,Kashani-Poor:2006si}. 
However, although we have some examples of these type of manifolds \cite{Villadoro:2005cu,House:2005yc,Camara:2005dc,Grana:2006kf} a 
general construction along the lines of CY manifolds is still missing. Therefore all that can be checked are general properties, in particular we do not 
have a full understanding of the moduli space of these manifolds. 

At this point we would like to clarify what we mean by moduli space. In a CY compactification without flux the space of massless 
deformations of the CY geometry forms the CY moduli space. Once flux is introduced some of the moduli pick up a mass and so strictly speaking they are no longer moduli. However they still correspond to deformations of the space that keep it CY. It is in this more general sense that  
we can define the moduli as fields that correspond to deformations of the space that keep it within its class of space. 
For example if we consider the mirror to a CY with electric H-flux then this would be a manifold with 
$SU(3)$ structure and we would call the moduli space the space of deformations that keep it within the class of manifolds with $SU(3)$-structure. 
This is a difficult space to define and study but some progress has been made \cite{Tomasiello:2005bp,House:2005yc,Grana:2005ny,Kashani-Poor:2006si,Grana:2006hr}.
We can conjecture that since mirror symmetry worked so exactly without flux it works equally well with flux and 
therefore the moduli space has the same structure as the moduli space of the mirror CY, i.e. that the four-dimensional theories are exactly the same.  

The two and three leg cases are termed magnetic fluxes and these lead to non-geometric compactifications. The question of what are the 
manifolds that lead to the mirror theories has recently been partially answered in \cite{Grana:2005ny,Benmachiche:2006df,Grana:2006hr,Micu:2007rd} 
where it was argued that they correspond to compactifications 
on manifolds with $SU(3)\times SU(3)$ structure. This is not yet fully understood but it seems that these manifolds still exhibit some similarities to 
CY compactifications. In this paper we therefore assume that indeed the four-dimensional theories arising from these compactifications are exactly 
mirror to the CY compactifications with H-flux. This allows us to study their superpotential using CY techniques. 

So far we have not suggested any alteration to the four-dimensional theory arising from type IIB compactifications on CY with H-flux. We have just 
suggested that the same theories can be derived from compactifications on more complicated geometries but without the presence of flux. However now 
we can construct new theories by turning back on normal fluxes in the complicated geometry compactifications. 
We assume turning them on is allowed, i.e. that they can be superposed on top of the non-geometric fluxes.   
The assumption relies on that fact that it seems there is no reason why turning them on should be forbidden:
The effective four-dimensional theories can be derived from generalised compactifications \cite{Benmachiche:2006df,Grana:2006hr,Micu:2007rd}.
It is consistent with supergravity gauging \cite{D'Auria:2007ay}. Also the newly introduced fluxes are simply symplectic 
rotations of the fluxes that are already present \cite{Tomasiello:2005bp,Grana:2006hr}. 
However we should keep in mind that turning them on takes us away from the realm of compactifications that correspond to a known solution of string theory.

%%%%%%%%%%%%%%%%%%%%%%%%%%%%%%%%%%%%%%%%%%%%%%%%%%%%%%%%%%%%%%%%%%%%%%%%%%%%%%%%%%%%%%%%%%%%%%%%%%%%%%%%%
\subsection{The four-dimensional superpotential}
\label{4dsuperpot}
%%%%%%%%%%%%%%%%%%%%%%%%%%%%%%%%%%%%%%%%%%%%%%%%%%%%%%%%%%%%%%%%%%%%%%%%%%%%%%%%%%%%%%%%%%%%%%%%%%%%%%%%%%

Consider compactifications of type IIB string theory on a CY orientifold (O3/O7) in the presence of fluxes \cite{Grimm:2004uq}. This results in an 
$\N=1$ theory with superfields whose scalar components are the moduli fields and axions. We keep the notation 
of moduli and axions even though, once we turn on more fluxes, the moduli are not in general flat directions and the axions will not in general have shift symmetries. In the Kahler sector we have the superfields 
\be
T_k = b_k + i t_k \;,
\ee
where the index $k$ runs over the even (under the orientifold action) four-forms $\tilde{\omega}_k \in H^{(2,2)} _{+}$ \cite{Grimm:2004uq}. 
The axions $b_k$ arise from the decomposition of the RR four-form $C_4$ and the $t_k$ are the Kahler moduli that come from the divisors of the CY. The complex-structure moduli form superfields in themselves which we label $U^a=z^a$, where the index $a=1,...,h_{-}^{(2,1)}$. From here on we suppress the orientifold parity on the hodge numbers. 

Following the introduction of three-form fluxes $F_3$ and $H_3$ a superpotential is induced for the complex-structure superfields.
The superpotential is given by \cite{Gukov:1999ya}
\be
W^{IIB} = \left( f_{\Lambda} - \phi h_{\Lambda} \right) \Pi^{\Lambda} \;. \label{wiibcy}
\ee
Here $f_{\Lambda}$, $h_{\Lambda}$ and $\Pi^{\Lambda}$ are all vectors with $2\left( h^{(2,1)}+1 \right)$ entries. We henceforth suppress contracted indices to save clutter.
The flux vector $f$ constitutes the RR fluxes and $h$ is the NS fluxes. They both have integer entries due to flux quantisation. The top $h^{(2,1)}+1$ entries of $h$ are the magnetic fluxes and the bottom half 
are the electric fluxes. We use the following index conventions to label the entries. The indices $\Lambda,\Sigma$ run 
over the full vector $\Lambda=0,...,2h^{(2,1)}+1$. These are then split into magnetic $\tA,\tB$ and electric $A,B$ indices that run over the range $0,...,h^{(2,1)}$ and label the top half and bottom half entries of the vector respectively. Finally there are the indices $\ta,\tb$ and $a,b$ which are the magnetic and electric indices with the $0$ component emitted $a,\ta=1,...,h^{(2,1)}$. Note that since they run over the same ranges it is possible to contract electric and magnetic indices. 
$\Pi$ is the period vector which is composed of the periods of the holomorphic three-form $\Omega$ 
\be
\Pi = \left( G_0, G_1, ..., G_{h^{(2,1)}},Z^0, Z^1, ...,Z^{h^{(2,1)}}\right) = \left(G_{\tilde{A}},Z^A\right)\;. \label{periodvec}
\ee
The $Z^A$ are the electric periods. They are homogeneous functions of the $h^{(2,1)}$ complex-structure moduli. The magnetic periods are functions of the electric ones $G_A\left( Z^A \right)$ and are given in terms of derivatives of a prepotential $G_A = \partial_A {\cal F}\left( Z^A \right)$. The prepotential is 
a homogeneous function of degree two which in the large commplex-structure limit takes the form 
\be
{\cal F}\left( Z^A \right) = \frac16 \kappa_{abc} \frac{Z^aZ^bZ^c}{Z^0} + \kappa^{(1)}_{ab}Z^aZ^b + \kappa^{(2)}_{a}Z^0Z^a + \left(Z^0\right)^2 \xi + \O\left( e^{iZ^a}\right) \;. \label{genprepot}
\ee
Here $\kappa_{abc}$ are integer constants, $\kappa^{(1)}_{ab}$ and $\kappa^{(2)}_{a}$ are rational, $\xi$ is a complex number.
The leading cubic term dominates in the large complex-structure limit in which case the superpotential becomes a polynomial in the superfields with rational coefficients. 
The other terms in the prepotential arise because of the complicated geometry of the complex-structure moduli space. 
Note that the superfields are given by $U^a = \frac{Z^a}{Z^0}$.

Apart from inducing a superpotential for the moduli, the fluxes also back-react on the geometry of the CY. In the IIB case this does not lead to a drastic 
change in the geometry as they just induces a warp factor so that the manifold is still conformally CY. In particular this back-reaction does not correct the superpotential \cite{DeWolfe:2002nn}. Note also that the Kahler moduli do not appear in the superpotential perturbatively, but may appear non-perturbatively due to E3 instantons 
or gaugino condensation on the world-volume of D7 branes wrapping four-cycles \cite{Becker:1995kb,Witten:1996bn,Harvey:1999as}.

Consider the mirror compactification to the IIB case discussed above. The presence of the flux vector $h$ implies that the mirror IIA compactification is not CY. Electric fluxes in $h$ imply that the mirror should be on a manifold with torsion and magnetic fluxes in $h$ imply that the mirror should 
also be non-geometric. These compactifications were studied in \cite{Benmachiche:2006df,Grana:2006hr,Micu:2007rd} where it was found that they should correspond to compactifications on orientifolds 
with $SU(3) \times SU(3)$ structure. The low-energy fields in such a configuration are the Kahler superfields $T^a$ composed of the Kahler form and the NS two-form, the complex-structure superfields $U_k$ composed of $\Omega$ and the RR three-form $C_3$, and the dilaton superfield $S$. 
The mirror superpotential in the large volume limit was derived in \cite{Micu:2007rd} and reads
\ba
W^{IIA} &=& -\frac{1}{6} f_0 \kappa_{abc} T^a T^b T^c + \half \kappa_{abc} f_{\ta} T^b T^c +  f_{a} T^a + f_{h^{(2,1)+1}} \nn \\
&\;& \;  - S \left( \frac{1}{2} \kappa_{abc} h_{\tilde{a}} T^b T^c + h_{a} T^a + h_{h^{(2,1)}+1} \right) \;. \label{wiiamirror}
\ea
Here $\kappa_{abc}$ denote the intersection numbers of the manifold. The various fluxes have been labelled according to their IIB mirrors, but they do not 
have the same origin as on the IIB side. The RR fluxes $f$ still come from the RR sector which is composed of the even RR forms. The flux $h_{h^{(2,1)}+1}$ \footnote{Note that the subscript $h^{(2,1)}$ refers to the Hodge numbers on the IIB side.} 
does have the same origin as the IIB case and comes from the NS three-form $H$. However the rest of the $h$ fluxes are geometric and non-geometric in 
origin. The fluxes $h_{a}$ are metric fluxes. The fluxes $h_{\ta}$ are non-geometric fluxes, denoted as $Q$-fluxes in the 
literature\footnote{Schematically the geometric and non-geometric fluxes correspond to changes in the derivative operator so that $D\Omega \sim h_{\ta} J + h_{a} J \w J$ \cite{Grana:2006hr,Micu:2007rd}.}. We have set $h_0=0$. The reason for this is that these non-geometric fluxes are highly non-geometric in that they 
correspond to manifolds that 
are not even locally geometric. They are referred to in the literature as $R$-fluxes and correspond to performing three T-dualities all along directions 
with H-flux. Throughout this paper we retain only non-geometric fluxes that correspond to manifolds that are locally geometric but only globally not so, these are the $Q$ fluxes \cite{Shelton:2005cf}. 

We are interested in comparing (\ref{wiiamirror}) with (\ref{wiibcy}). We see that we should identify the superfields 
\be
U^a \leftrightarrow T^a \;,\; \phi \leftrightarrow S \;.
\ee
This corresponds to the interchange of the complex-structure and Kahler moduli. 
We see that the two superpotentials match if we take the large complex-structure limit on the IIB side. However, if mirror symmetry is to hold fully 
they should match at all points in the complex-structure moduli space. Consider turning off the geometric and non-geometric $h$ fluxes so that on both sides 
of the mirror we have a CY manifold. Then mirror symmetry should be fully implemented by including perturbative ($\alpha'$) and non-perturbative (world-sheet instanton) to the IIA superpotential. If this also holds in the presence of the $h$ flux, then it implies that if we know the full geometry of the complex-structure moduli space of the CY on the IIB side then we know all these corrections to the superpotential on the IIA side. This is highly non-trivial since the manifold we are compactifying on is non-geometric. We have already performed a simple check of such an assumption by matching the superpotential in the large volume large complex-structure limit. 

On the type IIA side we have yet to turn on NS fluxes, which we denote $e_{h^{(2,1)}+1}^{k}$. Turning them on induces a term in the superpotential 
\be
W_{IIA}^{'} = -e_{h^{(2,1)}+1}^k U_k \label{iiansw}\;.
\ee
If we now employ mirror symmetry we would reach a type IIB configuration which would have a similar term but with the Kahler superfields, $T_k$. 
The origin of such a term is non-geometric and so we see that, as expected, the IIA NS fluxes map to non-geometric fluxes on the IIB side\footnote{Schematically they correspond to a change in the derivative operator so that $D(J\w J) \sim e_{h^{(2,1)}+1}^k \Omega $}.
The NS fluxes $e_{h^{(2,1)}+1}^k$ form a row in a matrix $e_{\Lambda}^k$ which now has the full symplectic index. 
The matrix is filled with the elements $e_a^k$, and $e_{\ta}^k$ ( we have set $e_0^k = 0$ ). On the IIA side the fluxes $e_a^k$ are metric fluxes which can 
be generated by a symplectic rotation of the present metric fluxes $h_{a}$. The fluxes $e_{\ta}^k$ are non-geometric Q-fluxes. 
On the IIB side both $e_a^k$ and $e_{\ta}^k$ are non-geometric $Q$-fluxes. 
The fluxes induce terms in IIA 
\be
W_{IIA}^{''} = -T^a e_a^{\;\;k} U_k - \half {\kappa}_{abc} T^b T^c e_{\ta}^{\;\;k} U_k \;. \label{iiametric}
\ee 
Now consider the possible small volume corrections to these terms, or equivalently the small complex-structure corrections to their type IIB mirrors. 
The terms look like they are the mirrors to a IIB large-complex structure limit of a term 
\be
W^{'+''}_{IIB} = - T\; e\; \Pi \;.
\ee
Indeed, in the large complex-structure limit, this is the term induced by the non-geometric fluxes on the IIB side \cite{Micu:2007rd}. 
Then it is reasonable to conjecture that the full corrections with respect to the IIA Kahler superfields 
are included in the full expression for the IIB complex-structure periods. More crucially, for our analysis in section \ref{susymink}, we consider the periods on the 
IIB side to be of the form of the CY periods. In that sense these models can be thought of as toy models. It could be that introducing the fluxes (\ref{iiansw}) and 
(\ref{iiametric}) changes the form of the periods in the superpotential. However we have already seen that mirror symmetry implies that even non-geometric compactifications 
still retain CY data in their superpotential. In the case that the form of the periods does change it is likely that similar considerations to those presented in section 
\ref{susymink} will hold. 

There may also be corrections to the superpotential in the IIA complex-structure fields, which are the IIB Kahler fields, i.e. corrections away from the IIB (IIA) large volume  (complex-structure) limit. However these fields can again be thought 
of as coming from a prepotential with electric and magnetic periods. The key point is that only one type of the periods, say magnetic, appears in the superpotential. Then 
the fields are exactly the magnetic periods, which is why they only appear linearly. The complicated form of the fully corrected prepotential 
would then show itself if we were to try to write the electric periods in terms of the superfields, but since these do not appear in the superpotential, any corrections to 
the prepotential do not alter the superpotential but only the Kahler potential\footnote{The explicit situation is quite complicated due to the presence of the orientifold. As shown in \cite{Grimm:2004ua}, indeed the superfields $U_k$ can be written as derivatives of a prepotential which takes a cubic form in the large complex-structure limit. In that sense they are like magnetic periods and would take a complicated form, away from the large complex-structure limit, 
if we were to try and write them in terms of the electric periods.}. This situation can be seen more clearly in a IIB setting where we only have electric NS fluxes and we 
move away from the large complex-structure limit. In that case although the prepotential takes a complicated form, the superpotential, which is only linear in the 
superfields $U_a$, is not affected by the corrections.

The metric fluxes $e_a^k$ do not have mirrors in IIB that are normal fluxes, i.e. non-vanishing values for field-strengths. 
With the exception of the fluxes $e_{\ta}^{k}$, which are not essential for our analysis and may be turned off if required, these are the only fluxes in our set up for which this is the case, all the other fluxes are either themselves normal fluxes or have mirrors that are. 
However it has been argued in \cite{Louis:2006kb} that they do have normal flux duals if we consider heterotic string compactifications. In those set-ups they would correspond to non-vanishing field-strengths for heterotic gauge fields. 

To summarise, in this paper we study the following type IIB superpotential
\be
W^{IIB} = \left( f - \phi\; h  - T \; e \right) \Pi \label{wiibgen} \;.
\ee
We assume that this is an exact expression up to non-perturbative effects in the Kahler moduli.
This is an assumption which we have tried to justify in this section but it may be that the superpotential receives further perturbative corrections in which case 
it would require further study to see if vacua that are exact with respect to these corrections can be found. 
We also have the fully corrected IIA mirror set-up which can be reached by mirror symmetry. The flux vectors $f$, $h$ and matrix $e$ have integer entries and we have set the fluxes that, on some side of the mirror, correspond to highly non-geometric $R$ fluxes to zero
\be
h_{0} = e_0^k = 0 \;.
\ee
The periods in (\ref{wiibgen}) are taken to be of the form of CY periods. 
The idea is then to use this to find exact Minkowski solutions to the supersymmetry equations arising from (\ref{wiibgen}). 
We study the existence of such solutions in the next section. The effects, and existence, of the possible non-perturbative corrections in the Kahler superfields are discussed in section \ref{nonper}. 

%%%%%%%%%%%%%%%%%%%%%%%%%%%%%%%%%%%%%%%%%%%%%%%%%%%%%%%%%%%%%%%%%%%%%%%%%%%%%%%%%%%%%%%%%%%%%%%%%%%%%%%%%
\subsection{The tadpoles}
%%%%%%%%%%%%%%%%%%%%%%%%%%%%%%%%%%%%%%%%%%%%%%%%%%%%%%%%%%%%%%%%%%%%%%%%%%%%%%%%%%%%%%%%%%%%%%%%%%%%%%%%%%

There are some constraints on the fluxes derived from the absence of tadpoles in the four-dimensional theory, or 
equivalently, from the ten-dimensional Bianchi identities. These were obtained in \cite{Grana:2006hr,Micu:2007rd} and read\footnote{Note that the last equation in (\ref{tadpoles}) can also pick up localised source contributions from NS5 branes and from KK monopoles as well as possible non-geometric sources \cite{Villadoro:2007yq}. We do not consider these in this paper.}
\ba
f\; \Sigma\; h = Q_0 \;\;,\; \left(f\;\Sigma\;e \right)^k = Q^k \;\;,\; \left(h\;\Sigma\;e \right)^k = 0 \;, \label{tadpoles}
\ea
where 
\be
\Sigma = \left( \begin{array}{cc} 0  & 1 \\ -1 & 0  \end{array}  \right) \;.
\ee
The local charges are given by 
\ba
IIB &:& Q_0 = 2 N^{D3} - \half N^{O3} \;\;,\; Q^k = 2 N^{k,D7} - 8 N^{k,O7} \;,\nn \\
IIA &:& Q_0 = 2 N_0^{D6} - 4 N_0^{O6} \;\;,\; Q^k = 2 N^{k,D6} - 4 N^{k,O6} \;,
\ea
where, for example, $N^{k,07}$ denotes the number of $O7$ planes wrapped on the four-cycle $\tilde{\omega}^k$.

%%%%%%%%%%%%%%%%%%%%%%%%%%%%%%%%%%%%%%%%%%%%%%%%%%%%%%%%%%%%%%%%%%%%%%%%%%%%%%%%%%%%%%%%%%%%%%%%%%%%%%%%%
\section{Supersymmetric Minkowski vacua}
\label{susymink}
%%%%%%%%%%%%%%%%%%%%%%%%%%%%%%%%%%%%%%%%%%%%%%%%%%%%%%%%%%%%%%%%%%%%%%%%%%%%%%%%%%%%%%%%%%%%%%%%%%%%%%%%%%

In this section we look for supersymmetric Minkowski vacua to the superpotential (\ref{wiibgen}). 
We use the full expression for the period vector $\Pi$ rather than taking the large complex-structure limit. For now we neglect possible non-perturbative corrections in the Kahler moduli.
We work primarily on the IIB side but the vacua equally apply on the IIA side under the mirror map. 
As discussed in section \ref{nongeocy}, we take the periods to be those of CY manifolds. They correspond to the periods 
of the original CY prior to performing one mirror symmetry, turning on more fluxes in IIA, then performing mirror symmetry again back to IIB. We note though that taking CY periods is primarily so that we can perform some 
explicit calculations and many of the ideas discussed in this section would hold equally well for a more general form of the periods. We look for solutions to the supersymmetry equations 
\be
\partial_{U^a}W = \partial_{T_k}W = \partial_{\phi}W = W = 0 \;,
\ee
where the superpotential is given by (\ref{wiibgen}). The equations can be written as
\ba
h \Pi &=& 0 \;, \label{fh} \\
f \Pi &=& 0 \;, \label{ff} \\
\left( e \Pi \right)^k &=& 0 \;, \label{fe}\\
\left( f - \phi h - e T \right) \partial_{a} \Pi &=& 0  \label{fu}\;.
\ea
These form $h^{(1,1)}+h^{(2,1)}+2$ complex equations and we have $h^{(1,1)}+h^{(2,1)}+1$ superfields and so in general there are no solutions.
However there may be solutions for special values of the fluxes. For example we may consider solving equations (\ref{ff}), (\ref{fe}) and (\ref{fu}), 
and then substituting the solutions into (\ref{fh}). This then becomes a constraint equation on the fluxes which we can try to pick so as to satisfy it. 
Since the fluxes are quantised, we are not guaranteed that a solution exists to this constraint. 
In the large complex-structure limit the periods are polynomials of the superfields with rational co-efficients, in which case it is possible to to solve the constraint 
and these are the solutions found in \cite{Micu:2007rd}.
However, in general at some point in moduli space, 
the periods would take transcendental values so that for no choice of fluxes can we cancel the terms in (\ref{fh}) against each other. 
Then each term must vanish by itself for which the only solution is for all the fluxes to vanish. It may be that for some special values 
in the complex-structure moduli space the periods take values that are not transcendental but rather lie in a finite extension over the rationals \cite{DeWolfe:2004ns}. 
This means that they can be written as some linear combination of irrational numbers with rational co-efficients. For example, the finite extension 
over the rationals given by numbers of the form 
\be
N = Q_1 + Q_2 \sqrt{3} \;,
\ee
where $Q_1$ and $Q_2$ are rational numbers, is denoted $\Q\left[\sqrt{3}\right]$. The dimension $\D$ of the extension is two. More generally we can think of the periods as taking values in some vector space $\Vol$ whose 
dimension is the dimension of the extension over the rationals $\mathrm{dim\;\Vol} = \D$. Then at general points in moduli space all the periods are linearly independent, with respect to rational coefficients, and the dimension of this vector space is $\D=b^3=2\left( h^{(2,1)}+1 \right)$. At particular points however the dimension may reduce. Then equation (\ref{fh}) becomes $\D<b^3$ constraint equations but now with rational coefficients multiplying the fluxes. These would in general have non-zero solutions since there are more fluxes than equations. 

We see that the question of how to solve the supersymmetry equations can be answered by looking for points in moduli space where the periods take values in a vector space which has dimension less than the number of fluxes in each flux vector (or row of the matrix $e$). We will see that there are also other constraints on the values which the periods can take. In order to understand how to work with the periods in such a way we need to recall their form. 

%%%%%%%%%%%%%%%%%%%%%%%%%%%%%%%%%%%%%%%%%%%%%%%%%%%%%%%%%%%%%%%%%%%%%%%%%%%%%%%%%%%%%%%%%%%%%%%%%%%%%%%%%
\subsection{The periods for Fermat CYs}
\label{perfer}
%%%%%%%%%%%%%%%%%%%%%%%%%%%%%%%%%%%%%%%%%%%%%%%%%%%%%%%%%%%%%%%%%%%%%%%%%%%%%%%%%%%%%%%%%%%%%%%%%%%%%%%%%%

We follow here the discussion of \cite{Candelas:1993dm,DeWolfe:2005gy}. We work with CYs\footnote{We assume that the orientifold projection does not project out the complex-structure moduli that we consider.} that are hyper-surfaces in weighted projective space, $WCP^4_{k_1k_2k_3k_4k_5}$, which has weighted homogeneous coordinates $x_i \sim \lambda^{k_i}x_i$ and degree $d=\sum_{i=1}^5k_i$. The CY is the vanishing locus 
of the polynomial 
\be
P(x_i)=x_1^{d/k_1} + x_2^{d/k_2} + x_3^{d/k_3} + x_4^{d/k_4} + x_5^{d/k_5} - d \psi x_1x_2x_3x_4x_5 + \sum_{a=1}^{h^{(2,1)}-1}\vp^a M_a(x_i)\;. \label{defpoly}
\ee
This form is valid only if all the $k_i$ divide $d$ in which case the CY is called Fermat. This will be the case for our examples and so, for the purposes of this paper, it suffices to consider this class only. The fields $\psi$ and $\vp^a$ are the complex-structure moduli 
of the manifold, and the $M_a(x_i)$ are monomials associated with each modulus. The periods for such a CY are constructed as follows. 
We define the fundamental period $\varpi_0$ through the integration of the holomorphic three-form over a cycle, which then defines our 
choice of cycle basis. The expression for the period reads
\be
\varpi_0 = -\sum_{n=1}^{\infty} \frac{\Gamma\left(\frac{n}{d}\right)\alpha^{n(d-1)/2}\left(d\psi\right)^{n-1}}{\Gamma\left(n\right)\Pi_{i=2}^5\Gamma\left(1-\frac{k_in}{d}\right)}U_n\left(\vp^a\right) \equiv \sum_{n=1}^{\infty} c_n U_n\left(\vp^a\right) \psi^{n-1} \;.
\ee
Here $\alpha$ is the $d^{\mathrm{th}}$ root of unity. The functions $U_n\left(\vp^a\right)$ are defined so that 
$U_n\left(\vp^a=0\right)=1$. We have taken $k_1=1$, to simplify things. Consider the polynomial 
(\ref{defpoly}), we can rotate the coordinate $x_1$ using $\alpha$ which would leave the first five terms invariant\footnote{The coordinate $x_1$ is only singled out for simplicity. More generally there is a monodromy group ${\mathbb Z}_d$ which we have chosen to represent through rotations on $x_1$.}. 
The monomials are not invariant but can be made so by an appropriate rotation of the moduli. This monodromy group $\A \sim {\mathbb Z}_d$ 
is a subgroup of the full monodromy group on the moduli space. It acts as 
\be
\A : \;\; \psi \rightarrow \alpha \psi \;,\;\; \vp^a \rightarrow \alpha^{Q_a} \vp^a \;,
\ee
where, for the case of $k_1=1$, the $Q_a$ are the powers of $x_1$ in the monomials. 
The fundamental period faithfully represents this group, and so we can generate $d-1$ new periods by acting on it with this monodromy 
group. However not all of these periods will be independent and, for Fermat models, there will be $b_3$ independent periods 
\be
\varpi_J = \sum_{n=1}^{\infty} c_n \alpha^{nJ}U_n\left(\alpha^{Q_aJ}\vp^a\right)\psi^{n-1} \equiv \sum_{n=1}^{\infty} c_n \left(p_n\right)_J\left(\vp^a\right)\psi^{n-1}\;,
\ee
where $J=0,...,b^3-1$. 
The coefficients $c_n$ are transcendental and may be irrational, and therefore if we want to find points where the periods 
are linearly related we should truncate the series in $n$ as much as possible\footnote{The series is not infinite since $c_n=0$ for 
$n=\frac{dl}{k_i}$ where $l$ is any positive integer, and for any $k_i$ with $i=2,3,4,5$.}. We can do this by setting $\psi$ to lie at the 
Landau-Ginzburg point
\be
\psi=0 \;,
\ee
which leaves only the $n=1$ term as non-vanishing. This simplifies the periods considerably and will allow us to find 
exact solutions. We therefore have 
\be
\varpi_J\left(\psi=0,\phi^a\right) = c_1 P_J\left(\phi^a\right) \;,
\ee
where we denote $\left(p_1\right)_J = P_J$.

The period vector (\ref{periodvec}) is obtained from the Picard-Fuchs basis $\varpi_J$ by an appropriate matrix $m$ with rational coefficients
\be
\Pi = m \varpi \;.
\ee
We can therefore write the supersymmetry equations at $\psi=0$ as
\ba
\bar{h} P &=& 0 \;, \label{fbh} \\
\bar{f} P &=& 0 \;, \label{fbf} \\
\left( \bar{e} P \right)^k &=& 0 \;, \label{fbe}\\
\left( \bar{f} - \phi \bar{h} - \bar{e} T \right) p_2 &=& 0  \label{fbupsi} \;, \\
\left( \bar{f} - \phi \bar{h} - \bar{e} T \right) \partial_{\vp^a} P &=& 0  \label{fbuphi} \;, 
\ea
where we have defined $\bar{h}=hm$, $\bar{f}=fm$, $\bar{e}=em$. It is these equations that we will attempt to solve in the 
upcoming sections for different toy models. These are toy models in the sense that the complex-structure moduli space 
will correspond to that of a CY manifold, and we will keep the number of Kahler moduli as free. This is for practical reasons since 
it is only the very simple cases with small numbers of complex-structure moduli that we can study in detail. There is no reason 
why the discussions should not hold for larger, more realistic, numbers of moduli. 

The only general constraint is that we can only perturbatively fix at most $h^{(2,1)}-1$ Kahler moduli. This follows from the fact that, since the dilaton and Kahler moduli only appear linearly in the superpotential, they are only constraint by $h^{(2,1)}$ equations \cite{Micu:2007rd}. 
If we have a larger number of Kahler moduli they can still be fixed but only non-perturbatively.

In the complex-structure moduli sector there are two types of cases; the case where there is only one complex-structure modulus $\psi$, and the case of multiple complex-structure moduli. We consider these cases separately in the upcoming sections.

%%%%%%%%%%%%%%%%%%%%%%%%%%%%%%%%%%%%%%%%%%%%%%%%%%%%%%%%%%%%%%%%%%%%%%%%%%%%%%%%%%%%%%%%%%%%%%%%%%%%%%%%%
\subsection{One complex-structure modulus}
%%%%%%%%%%%%%%%%%%%%%%%%%%%%%%%%%%%%%%%%%%%%%%%%%%%%%%%%%%%%%%%%%%%%%%%%%%%%%%%%%%%%%%%%%%%%%%%%%%%%%%%%%%

In this case all the non-geometric fluxes are switched off on the IIB side and we return to a CY manifold. On the 
IIA side however we still have both metric and non-geometric fluxes and so this provides a non-trivial example of 
moduli stabilisation in IIA. Although the manifold is more complicated on the IIA side there is a conceptual advantage, 
this is because, with the exception of some exotic Gepner models \cite{Becker:2006ks,Becker:2007dn}, such a setup could never fix all the moduli on the IIB side as there must be at least one Kahler modulus that measures the volume. 
On the IIA side we simply have no complex-structure moduli, this is a situation that could easily arise since that are many known examples with a 
rigid complex-structure (all nearly Kahler manifolds are for example \cite{Micu:2004tz}).

This case was studied on the CY IIB side in \cite{DeWolfe:2004ns}, we review it here as a warm-up example. 
In this case the periods evaluated at the 
Landau-Ginzburg point become powers of $\alpha$. This means that they are element of the cyclotomic group $\F_d$ which 
is the group generated by the $d^{\mathrm{th}}$ roots of unity. Recall that we are interested in how many of the periods are independent. The cyclotomic group $\F_d$ has $\phi(d)$ independent elements where $\phi(d)$ is the Euler totient function which is given by the number of integers less than $d$ that are relatively prime to $d$. 
Hence, the periods lie in a vector space $\Vol$ of dimension $\D=\phi(d)$.
In order to solve the equations (\ref{fbh}) and (\ref{fbf}) with non-vanishing fluxes we therefore require that $b^3=4>\phi(d)$. 
This is not satisfied for all the one complex-structure modulus CYs, but holds for a substantial fraction of them. 
Once the equations (\ref{fbh}) and (\ref{fbf}) are solved for some choice of fluxes, the remaining equation (\ref{fbupsi}), which now has $\bar{e}=0$, can be solved for the dilaton. Finally we should impose the single tadpole 
equation, which is the first equation in (\ref{tadpoles}), on the remaining $8-2\phi(d)$ free fluxes. 

As an example of such a construction consider the CY manifold, $M_6$ in \cite{DeWolfe:2004ns}, which is given by the polynomial\footnote{Actually the CY $M_6$ also has a large number of monomials with other complex-structure moduli, however, since it is 
Fermat the mirror polynomial just has all the monomials that are not invariant under the symmetries of the fundamental 
monomial (the monomial multiplying $\psi$) projected out. Therefore the mirror has one complex-structure modulus and is 
the actual manifold we work with, however its polynomial is the same as the original polynomial with all the monomials projected out and so we can just use (\ref{cymon1}). See for example \cite{DeWolfe:2004ns} for discussions on this point.}
\be
2x^3_1 +\sum_{i=2}^5 x_i^6  - 6 \psi x_1x_2x_3x_4x_5 = 0 \;\;,\; x \in WP^4_{[2,1,1,1,1]} \;. \label{cymon1}
\ee
We have $d=6$ which gives $\D=\phi(6)=2<4$ and so we expect non trivial solutions to the flux equations. Equation (\ref{fbf}) reads 
\be
\rf_0 + \rf_1 \alpha + \rf_2 \alpha^2 + \rf_3 \alpha^3 = 0 \;. \label{falpha1}
\ee
Now $\alpha=e^{i\pi/3}=\half \left( 1+i\sqrt{3}\right)$ which means that $\alpha^2=\alpha-1$. This is precisely the 
relation which reduced the $6$ elements in $\F_6$ to only $\phi(6)=2$ independent elements. Which means that (\ref{falpha1}) reads
\be
\left( \rf_0 - \rf_2 - \rf_3 \right) +\alpha\left( \rf_1 + \rf_2 \right) = 0 \;.
\ee
This can be solved by constraining $\rf_0=\rf_2+\rf_3$ and $\rf_1=-\rf_2$, leaving two free fluxes. The same equations 
hold for the $\rh$ fluxes. The last supersymmetry equation is solved for 
\be
\phi = \frac{\rf p_2}{\rh p_2}\;,
\ee
which gives
\be
\im{\phi} = \frac{\half\sqrt{3}\left[\rh_2\left(\rf_2+\rf_3\right)-\rf_2\left(\rh_2+\rh_3\right) \right]}{\rh_3^2 + \rh_2^2 + \rh_2\rh_3} \;.
\ee
Finally we need to impose the tadpole constraint, and also for the mirror configuration to remain locally 
geometric, i.e. with no $R$ fluxes, we require $h_0=0$. To impose these conditions we need to relate the two 
flux bases which is done be the explicit expression for the matrix $m$ derived in \cite{Klemm:1992tx}
\be
\left(  \rf_0 , \rf_1 , \rf_2 , \rf_3  \right) = \left( f_0 , f_1 , f_2 , f_3 \right) \left( \begin{array}{cccc} -\frac13  & -\frac13 & \frac13 & \frac13 \\ 0 & 0 & -1 & 0 \\ -1 & 0 & 3 & 2 \\ 0 & 1 & -1 & 0 \end{array}  \right) \;.
\ee
The supersymmetry constraints on the fluxes becomes $f_1=3f_2$ and $f_0=-3f_2-f_3$, with the same equations holding for $h$. In order to impose $h_0=0$ we further impose $h_3=-3h_2$.
With all these constraints the tadpole equation reads
\be
-2h_2 \left( f_3 - 3f_2 \right) = Q_0 \;.
\ee
We must choose our fluxes to satisfy this constraint. This has an important consequence for the dilaton which, after imposing the tadpoles, reads 
\be
\im{\phi} = \frac{-Q_0}{28\sqrt{3}\left(h_2\right)^2} \;.
\ee
We therefore see that the dilaton is capped by the orientifold charge and so its value is not parametrically controlled. 

%%%%%%%%%%%%%%%%%%%%%%%%%%%%%%%%%%%%%%%%%%%%%%%%%%%%%%%%%%%%%%%%%%%%%%%%%%%%%%%%%%%%%%%%%%%%%%%%%%%%%%%%%
\subsubsection{Discrete symmetries}
%%%%%%%%%%%%%%%%%%%%%%%%%%%%%%%%%%%%%%%%%%%%%%%%%%%%%%%%%%%%%%%%%%%%%%%%%%%%%%%%%%%%%%%%%%%%%%%%%%%%%%%%%%

The vanishing of the superpotential can sometimes be attributed to a discrete R symmetry of the moduli fields \cite{Dine:2004is,DeWolfe:2004ns,DeWolfe:2005gy,Dine:2005gz}. 
By this we mean a symmetry that acts on the fields, but not the fluxes, and leaves the superpotential invariant up to a phase rotation. 
Then if such an R symmetry exists the superpotential vanishes at values of the moduli that are at fixed points of the symmetry. 
In particular we look for R symmetries that are subgroups of the monodromy 
group, which acts on the periods as 
\be
A \Pi\left(\psi\right) = \alpha\Pi\left(\alpha\psi\right) \;,
\ee
where $A$ is a matrix representing $\A$. Then consider picking the fluxes so that the flux vectors are left eigenvectors, with eigenvalues unity, 
of some power $N<d$ of the monodromy matrix\footnote{The matrix $A$ has only complex eigenvalues and so only powers of it can have eigenvalues of unity. This is why, for general values of the dilaton,  
the R symmetry can only be a subgroup of the full monodromy group. It may be possible to enhance it to the full monodromy group at special values of the dilaton \cite{DeWolfe:2004ns,DeWolfe:2005gy}.} 
\be
g A^N = g \;,\;\; f A^N = f \;.
\ee
Using (\ref{wiibgen}) we see that 
\be
W\left(\alpha^N \psi\right) = \bar{\alpha}^N W \left(  \psi \right) \;.
\ee
This is an R symmetry, and the LG point is a fixed point of it which forces the superpotential to vanish. 
It is possible to check that indeed this is the mechanism at work in the case studied above. 
The supersymmetry equations are solved precisely for flux vectors that are left eigenvectors for $N=4$ \cite{DeWolfe:2004ns}.

This mechanism of forcing the superpotential to vanish through a discrete symmetry is a powerful one and we shall 
see can also be applied to the case where there are more than one complex-structure moduli, to which we now turn.

%%%%%%%%%%%%%%%%%%%%%%%%%%%%%%%%%%%%%%%%%%%%%%%%%%%%%%%%%%%%%%%%%%%%%%%%%%%%%%%%%%%%%%%%%%%%%%%%%%%%%%%%%
\subsection{Multiple complex-structure moduli}
\label{multigeo}
%%%%%%%%%%%%%%%%%%%%%%%%%%%%%%%%%%%%%%%%%%%%%%%%%%%%%%%%%%%%%%%%%%%%%%%%%%%%%%%%%%%%%%%%%%%%%%%%%%%%%%%%%%

In this section we study the more general situation where there are multiple complex-structure moduli. 
It is here that the new terms in the superpotential play a crucial role. To see this consider turning off all the 
non-geometric fluxes $e=0$. Then the Kahler moduli remain as perturbatively flat directions. 
We now concentrate on just the complex-structure moduli sector. Recall that to solve 
the equations (\ref{fbh}) and (\ref{fbf}) we require that the periods lie in a vector space of dimension less than $b^3$. As pointed out in \cite{DeWolfe:2005gy}, 
for the case of multiple complex-structure moduli, this can occur either 
at a some point in moduli space, as was the case for the one parameter models, or it can occur within a locus where 
the $\vp^a$ can take any values. We now also have extra equations to solve which are the F-terms for the 
complex-structure moduli (\ref{fbuphi}). For each modulus, that is a value for $a$, the derivatives of the periods 
$\partial_{\vp^a}P$ will form, like the periods, a vector space $\Vol^a$. Again, for a solution with non-vanishing fluxes, we require the dimension of this vector space to decrease below $b^3$. Hence to solve the full set of equations we require all the vector spaces $\Vol$ and $\Vol^a$ to reduce in dimension at the same time. This is automatic 
for the case where $\vp^a$ can take any values, as we outline below. In the case of isolated vacua generically this is 
not the case. However at the point $\vp^a=0$ indeed all the vector spaces reduce to the dimension of the extension of 
the cyclotomic fields. We study this possibility in more detail in section \ref{specsol}.

We now review the proposition in \cite{DeWolfe:2005gy} of how to construct solutions valid for all values of $\vp^a$. 
The idea is that in many 
models the dimension of $\Vol$ reduces for all values of the $\vp^a$. To see this consider the periods $P_J$ at the LG point $\psi=0$
\be
\left\{ P_0\left(\vp^a\right),P_1\left(\vp^a\right),...,P_{b^3-1}\left(\vp^a\right) \right\} = \left\{ P_0\left(\vp^a\right),\alpha P_0\left(\alpha^{Q_a}\vp^a\right),...,\alpha^{b^3-1}P_0\left(\alpha^{(b^3-1)Q_a}\vp^a\right) \right\}\;,
\ee
Now consider some integer $N<d$ which satisfies the periodicity condition
\be
\alpha^{NQ_a}=1 \;\; \forall a \;. \label{periodcon}
\ee 
Then all the periods with indices 
$J$ that are multiples of $N$ become $\alpha^JP_0\left(\vp^a\right)$ and so all lie, up to the overall multiplicative 
factor of $P_0\left(\vp^a\right)$, within $\F_d$ and so there can only be at most $\phi(d)$ independent ones within this set. Similarly for the rest of the periods, and the period set becomes periodic in $N$. Therefore a sufficient, 
but not necessary, condition for the dimension of $\Vol$ to reduce is that $b^3>N\phi(d)$. In general this condition 
is too strict since not all the elements of $\F_d$ need appear in each periodic set. Upon a reduction in the dimension 
of the vector space non-trivial flux solutions exist. 
Note that such a flux solution holds for all values of $\vp^a$. Now we are left with solving the equations 
involving $\partial_{\vp^a}P$ (\ref{fbuphi}), but the arguments above hold equally for the vector space $\Vol^a$ since the derivatives do not change the periodicities. All of the $\Vol^a$ also reduce in dimension and we can solve them as well. In fact they are automatically solved since, for this construction, the elements of the $\Vol^a$ obey the same linear relationships as the elements of $\Vol$.  

As we saw in the one parameter case it is possible to solve the equations (\ref{fbh}) and (\ref{fbf}) by looking 
for fluxes that satisfy $g=gA^N$. If this $N$ also satisfies the periodicity condition (\ref{periodcon}) then  
\be
g P\left(\vp^a\right)=\alpha^NgP\left(\vp^a\right) \;,
\ee
which implies that $g P=0$. This is a discrete R symmetry which imposes the vanishing of the superpotential. The important point here is that all values of $\vp^a$ are 
fixed points of the symmetry and so the superpotential vanishes for all values of the $\vp^a$.

Since $A^d=1$ the eigenvalues of $A$ are roots of unity. Then for each eigenvalue $a_k$ which satisfies the condition
\be
\left(a_k\right)^N=1 \;,
\ee
we have an appropriate flux eigenvector which solves the supersymmetry equations and leads to 
an R symmetry as above. These eigenvalues are the unfaithful eigenvalues of $A$ and their number can be determined 
as outlined in \cite{DeWolfe:2005gy}.  

To summarise, it was shown in \cite{DeWolfe:2005gy}, that each unfaithful eigenvalue of the monodromy matrix $A$ represents a flux eigenvector which is a solution to (\ref{fbh}) and (\ref{fbf}), and if it also satisfies the periodicity condition equation (\ref{fbuphi}) is automatically solved. Equation (\ref{fbupsi}) is solved by the dilaton. 

The solutions discussed above have, by construction, flat directions since they are valid for all values of $\vp^a$. 
They also have all the Kahler moduli perturbatively flat. 
For the rest of this section we analyse how this situation changes when we turn the non-geometric fluxes $e$ back on. 
Two fundamental changes occur following this generalisation. The first is that the Kahler moduli also appear in the supersymmetry 
equations and may get fixed perturbatively. The second is that we can use the Kahler moduli to solve some of the equations (\ref{fbuphi}). In that sense they play the role of 
the dilaton in the one complex-structure modulus case. Recall that it is these equations that forced us to look for solutions that are valid 
for all values of the $\vp^a$ and so with this constraint relaxed we can look for solutions that are only valid 
at isolated points in the complex-structure moduli space. Therefore we see that the two changes imply that we can 
find scenarios where all the moduli are fixed. 
The possible solutions to the equations (\ref{fbh}-\ref{fbuphi}) depend on the number of Kahler moduli present.
We now discuss the two key scenarios. 

%%%%%%%%%%%%%%%%%%%%%%%%%%%%%%%%%%%%%%%%%%%%%%%%%%%%%%%%%%%%%%%%%%%%%%%%%%%%%%%%%%%%%%%%%%%%%%%%%%%%%%%%%
\subsubsection{Direct Solutions :  $h^{(1,1)} \ge h^{(2,1)}-1$}
\label{dirsol}
%%%%%%%%%%%%%%%%%%%%%%%%%%%%%%%%%%%%%%%%%%%%%%%%%%%%%%%%%%%%%%%%%%%%%%%%%%%%%%%%%%%%%%%%%%%%%%%%%%%%%%%%%%

In this scenario we have enough Kahler moduli to completely solve the $h^{(2,1)}-1$ equations (\ref{fbuphi}). 
If the inequality is saturated then this fixes all the Kahler moduli perturbatively, otherwise they must be fixed non-perturbatively. 
We can use the dilaton to solve equation (\ref{fbupsi}) leaving the $h^{(1,1)}+2$ equations 
(\ref{fbh}-\ref{fbe}). We want to find solutions to these equations that are only valid 
at one point in the complex-structure moduli space. 
Consider equation (\ref{fbh}). We may attempt to solve this equation directly 
for the $\vp^a$ in terms of the fluxes. In general it is not guaranteed that such a solution exists within a physical 
regime for the $\vp^a$. However we will soon outline a construction where there is such a solution. If we grant that 
we have a solution, valid only at some point in moduli space, then at that point it must be that the dimension of $\Vol$ is reduced. This means that we can choose the fluxes $\bar{e}$ and $\bar{f}$ so that equations (\ref{fbf}) and (\ref{fbe}) are also satisfied at this point. In general at this point $\partial_{\vp^a}P\neq 0$ which means that the Kahler moduli 
all get fixed and so does the dilaton. Hence this leads to a scenario where all the moduli are fixed. There remain the tadpole equations to impose but these can always be satisfied as long as there are enough free fluxes left. 

We now outline how such a mechanism can work with a toy example. The example we consider is as follows.  
We take the complex-structure moduli space of the CY manifold $P_{[1,1,2,2,2]}$ studied in \cite{Candelas:1993dm}. This manifold 
has two complex-structure moduli $\psi$ and $\vp$. The manifold has $d=8$, which gives $\phi(8)=4$, and $Q_1=4$.
We also consider one Kahler modulus $T$. We can therefore solve 
directly equations (\ref{fbupsi}) and (\ref{fbuphi}) for $\phi$ and $T$
\ba
\phi &=& \frac{\bar{f}_{\psi}\bar{e}_{\vp}-\bar{e}_{\psi}\bar{f}_{\vp}}{\bar{e}_{\vp}\bar{h}_{\psi} - \bar{h}_{\vp}\bar{e}_{\psi}} \;, \label{twomodphi} \\ 
T &=& \frac{\bar{f}_{\vp}\bar{h}_{\psi}-\bar{h}_{\vp}\bar{f}_{\psi}}{\bar{e}_{\vp}\bar{h}_{\psi} - \bar{h}_{\vp}\bar{e}_{\psi}} \;, \label{twomodT}
\ea
where we use the notation $\bar{f}_{\vp}=\bar{f} \partial_{\vp}P$ and $\bar{f}_{\psi}=\bar{f} p_2$. Note that the matrix $\bar{e}$ is now a vector since $k$ only takes one value. 
The tadpoles require that $\bar{e}$, $\bar{f}$ and $\bar{h}$ must not be aligned. To solve the remaining equations consider equation (\ref{fbf})
\be
\bar{f} P\left( \vp \right) = \bar{f}_0 P_0\left( \vp \right) + \bar{f}_1 \alpha \tilde{P}_0\left( \vp \right) + \bar{f}_2 \alpha^2 P_0\left( \vp \right) + \bar{f}_3 \alpha^3 \tilde{P}_0\left( \vp \right) + \bar{f}_4 \alpha^4 P_0\left( \vp \right) + \bar{f}_5 \alpha^5 \tilde{P}_0\left( \vp \right) = 0\;,
\ee
where $\tilde{P}_0\left(\vp\right) = P_0\left(-\vp\right)$. This is solved for 
\be
\frac{\alpha\tilde{P}_0\left( \vp \right)}{P_0\left( \vp \right)} = \frac{\bar{f}_4 - \bar{f}_0 - i\bar{f}_2 }{\bar{f}_1 - \bar{f}_5 + i\bar{f}_3}\;. \label{moorevpeq}
\ee
Therefore the periods take values in the extension over the rationals $\Q[i]$ which is of dimension two. In \cite{Moore:1998pn} it was shown that (\ref{moorevpeq}) has solutions with $\vp$ at physical values for some choices of fluxes. The solution reads 
\be
\vp^2 = \frac{f^{24}(\tau)}{2^8} + \frac{2^8}{f^{24}(\tau)} + \half \;,
\ee
where $f(\tau)$ is the Weber function and $\tau$ is defined as 
\be
\tau = \left(1+i\right)\frac{\bar{f}_4 - \bar{f}_0 - i\bar{f}_2 }{\bar{f}_1 - \bar{f}_5 + i\bar{f}_3} -1 \;.
\ee
The solution for $\vp$ also solves equations (\ref{fbh}) and (\ref{fbe}) as long as 
\be
\frac{\bar{f}_4 - \bar{f}_0 - i\bar{f}_2 }{\bar{f}_1 - \bar{f}_5 + i\bar{f}_3} = \frac{\bar{h}_4 - \bar{h}_0 - i\bar{h}_2 }{\bar{h}_1 - \bar{h}_5 + i\bar{h}_3} = \frac{\bar{e}_4 - \bar{e}_0 - i\bar{e}_2 }{\bar{e}_1 - \bar{e}_5 + i\bar{e}_3} \;.
\ee
This places four real constraints on the flux vectors $\bar{h}$ and $\bar{e}$ as expected since the dimension of $\Vol$ has been reduced to two. We therefore have $18-4-2=12$ free fluxes at our disposal, where we have set $h_0=e_0=0$ to keep only $Q$ fluxes. These should be chosen to solve the three tadpole equations, and so that $\vp$, $T$ and $\phi$ take values in a physical regime. 

It is difficult to solve explicitly for the moduli in terms of the fluxes, as we did in the one modulus case, 
since it is not clear how to eliminate the periods in the expressions (\ref{twomodphi}) and (\ref{twomodT}).
This implies that we can not check if the solutions are such that we have parametric control over all the moduli. In the large complex-structure 
limit it was shown in  \cite{Micu:2007rd} that indeed all the moduli can be parametrically controlled. However the solutions were controlled by one parameter. This means 
that they can not apply to the above models since we have fixed one of the moduli to lie at the LG point and this fixes the controlling parameter to a small value. 
The relevant question here is can we keep one modulus small and parametrically control the others? Since we are unable to solve explicitly for the dilaton and 
Kahler modulus we can not fully answer this question. 

For the rest of this section we show a much weaker result which is that keeping one complex-structure modulus at the LG point the other one can be parametrically 
controlled. We should keep in mind that it may be that taking the complex-structure modulus large drives the dilaton and Kahler modulus to small values\footnote{A hint that this may not be the case is that in the solutions of \cite{Micu:2007rd} it is possible to choose the fluxes such that one of the complex-structure moduli is fixed at the LG point, 
whilst the other one is parametrically controlled. In that case for large values of the controlled complex-structure modulus the Kahler modulus tends to large values and the 
dilaton tends to a constant value. However this is only a hint since there are certainly corrections to that model as one of the complex-structure moduli is small.}.
We first solve generally for the complex-structure moduli. The basis conversion matrix for this case is given by \cite{Candelas:1993dm}
\be
m = \left( \begin{array}{cccccc}  -1 & 1 & 0 & 0 & 0 & 0   \\ 1 & 0 & 1 & -1 & 0 & -1   \\ \frac32 & 0 & 0 & 0 & -\half & 0   \\ 1 & 0 & 0 & 0 & 0 & 0 \\ -\frac14 & 0 & \half & 0 & \quarter & 0  \\ \quarter & \frac34 & -\half & \half & -\quarter & \quarter  \end{array} \right) \;.
\ee
This gives 
\ba
U^1 &=& \frac{Z^1}{Z^0} = -\quarter + \half \alpha^2 + \half \alpha^4 \;, \nn \\
U^2 &=& \frac{Z^2}{Z^0} = \left( \quarter -\half \alpha^2 -\quarter \alpha^4 \right) + \frac{\alpha\varpi_0\left(-\vp\right)}{\varpi_0\left(\vp\right)} \left(\frac34 + \half \alpha^2 + \quarter \alpha^4 \right) \;.
\ea
Which in the vacuum gives 
\ba
\im{U}^1 &=& \half \;,  \\
\im{U}^2 &=& -\half + \frac{\half}{\left( \bar{f}_1 - \bar{f}_5 \right)^2+\bar{f}_3^2} \left[ \bar{f}_2\left(\bar{f}_1 - \bar{f}_5 \right) - \bar{f}_3\left( \bar{f}_4 - \bar{f}_0 \right) +\left( \bar{f}_4 - \bar{f}_0 \right)\left( \bar{f}_1 - \bar{f}_5 \right) - \bar{f}_2\bar{f}_3   \right] \;. \nn
\ea
Note that one is fixed to a small value corresponding to taking $\psi=0$.

Now we must choose the fluxes to solve the supersymmetry equations (\ref{fbh}-\ref{fbuphi}) and the tadpoles (\ref{tadpoles}). The most general solution for the fluxes is very complicated. 
However, for the purposes of showing parametric control we can look for a simpler particular solution. This is still too complicated to display here and so we just outline the flux choices. We take
\be
\bar{e}_2 = \bar{e}_3 = \bar{e}_5 = \bar{h}_2 = \bar{h}_3  = \bar{f}_1  = 0 \;,\;\; \bar{f}_2 = \frac{-Q_0 + 7 \bar{f}_3 \bar{h}_5}{4 \bar{h}_5} \;. \label{paraflux}
\ee
This leaves eleven free fluxes and ensures $e_0=h_0=0$. Six are used to solve the three tadpole equations and the three (real) supersymmetry equations (one equation is satisfied identically for the choice of fluxes (\ref{paraflux})). This leaves five free fluxes which in our case are $\bar{e}_4, \bar{f}_3, \bar{f}_5, \bar{h}_4, \bar{h}_5$. 
In terms of these we find 
\ba
\im{U^1} &=& \half \;, \nn \\
\im{U^2} &=& -\half + \frac{\bar{e}_4\left(2\bar{f}_3\bar{h}_5 + Q_0 \right) - \bar{h}_4 Q_1 }{3\bar{h}_4Q_1} \;. 
\ea
The solutions lead to the large complex-structure limit for large $\bar{e}_4, \bar{f}_3, \bar{h}_5$. It can be checked that taking these large does not imply 
that any of the fluxes, fixed by the tadpoles and supersymmetry equations, must be less than one and therefore they could take integer values. 

In section \ref{multigeo} we saw that the vanishing of the superpotential in the vacuum can be associated with the 
presence of an R-symmetry that is the action of a subgroup of the monodromy group. In the class of solutions discussed in this section this is generally not the case. For the example studied, the monodromy group acts as $\vp \ra -\vp$, which for general fluxes is 
not an R-symmetry of the superpotential. For some choice of fluxes, $\bar{f}A^2=\bar{f}$ or $\bar{f}A^4=\bar{f}$, it leads to the type of R-symmetry discussed in section \ref{multigeo}.

%%%%%%%%%%%%%%%%%%%%%%%%%%%%%%%%%%%%%%%%%%%%%%%%%%%%%%%%%%%%%%%%%%%%%%%%%%%%%%%%%%%%%%%%%%%%%%%%%%%%%%%%%%%%%%%%%%%%%%%%%%
\subsubsection{Special solutions : $h^{(1,1)} \ge \;$ Number of complex structure moduli for which $\alpha^{(Q_a+1)N}=1$. }
\label{specsol}
%%%%%%%%%%%%%%%%%%%%%%%%%%%%%%%%%%%%%%%%%%%%%%%%%%%%%%%%%%%%%%%%%%%%%%%%%%%%%%%%%%%%%%%%%%%%%%%%%%%%%%%%%%%%%%%%%%%%%%%%%%

Solutions to (\ref{fbh}-\ref{fbe}) can also be obtained by fixing the value of the complex-structure fields to a point 
where the dimension of $\Vol$ and of the $\Vol^a$ all reduce simultaneously, and then choosing the fluxes 
to solve the resulting constraint equations. It allows greater freedom in the number of Kahler moduli required 
to solve all the equations, but, as discussed in section \ref{nongeoland}, is statistically less likely than the direct solution of section \ref{dirsol}.

At a point in moduli space where the dimension of $\Vol$ reduces, the dimensions of the $\Vol^a$ will generally not reduce. However at the point $\vp^a=0$, which we henceforth denote the Generalised LG (GLG) point, the dimensions all reduce to that of the extension over the rationals by the cyclotomic fields. 
We therefore limit the work in this section to that case whilst bearing in mind that there may also be other points in moduli space where such a universal dimensional reduction occurs. 

Recall that flux vectors that are left eigenvectors of some power of the monodromy matrix $A^N$ solve the 
supersymmetry equations (\ref{fbh}-\ref{fbe}) and lead to a discrete $R$ symmetry in the superpotential which forces it to vanishes at the fixed points. The number of independent such vectors is given by the number of unfaithful eigenvalues of $A$. If the power $N$ also satisfies the periodicity condition (\ref{periodcon}) then the supersymmetry 
equations (\ref{fbuphi}) are also automatically solved and the solutions are valid over all values of $\vp^a$. In order 
to find solutions that are only valid at the GLG point we require unfaithful eigenvalues of $A$ that do not satisfy the 
periodicity condition. This follows since 
\be
f P\left( \vp^a \right) = f A^N P \left( \vp^a \right) = \alpha^N f P \left( \alpha^{NQ_a}\vp^a \right) \;. \label{glgr}
\ee
At the GLG point this enforces $f P = 0$, but not away from it. This constraint fixes all the moduli that have 
a $Q_a$ which does not satisfy the periodicity condition (\ref{periodcon}). The rest are left as flat directions.  

Now we turn to the equations (\ref{fbuphi}). To see if they are solved we can act again with $A^N$ which gives 
\be
f \partial_{\vp^a} P\left( \vp^a \right) = f A^N \partial_{\vp^a} P\left( \vp^a \right) = \alpha^N \alpha^{Q_aN}f \partial_{\left(\alpha^{Q_aN}\vp^a\right)} P\left( \alpha^{Q_aN} \vp^a\right) \;. 
\ee
At the GLG point this vanishes as long as 
\be
\alpha^{\left( Q_a + 1 \right)N} \neq 1 \;. \label{fbphivanish}
\ee
Hence all the equations that satisfy (\ref{fbphivanish}) vanish automatically. The remaining equations can then 
be used to fix the Kahler moduli. It would be interesting to find CYs for which (\ref{fbphivanish}) is satisfied for all $a$ as these would lead to fully geometric 
isolated Minkowski vacua.

The CY $P_{[1,2,3,3,3]}$ is an example that has unfaithful eigenvalues that do not satisfy the 
periodicity condition. The CY has three complex-structure moduli. It has $d=12$, $Q_1=4$ and $Q_2=8$. The period matrix has four unfaithful eigenvalues $\alpha^2$, $\alpha^3$, $\alpha^9$ and $\alpha^{10}$. The 
eigenvalues $\alpha^3$ and $\alpha^9$ are for $N=4$. But since $N Q_1 = 16$ and $N Q_2 = 32$ we see that they do not 
satisfy the periodicity condition. We also have $N\left( Q_1 + 1\right)=20$ and $N\left( Q_2 + 1\right)=36$ which implies that $f \partial_{\vp^1}P = 0$ and $f \partial_{\vp^2}P \neq 0$. We therefore require one Kahler modulus in order 
to solve the non-vanishing equation. Lets see how this is realised explicitly.
The equation (\ref{fbf}) reads
\ba
&\;&\left( \bar{f}_0 - \bar{f}_6 + \alpha^3 \bar{f}_3 \right) P\left( \vp^1, \vp^2\right) + \left( -\bar{f}_4 + \left(\bar{f}_1-\bar{f}_7\right)\alpha + \bar{f}_4 \alpha^2 \right) P\left( (\alpha^2-1)\vp^1, -\alpha^2 \vp^2\right) \nn \\ &\;& + \left( -\bar{f}_5 \alpha + \bar{f}_2 \alpha^2 + \bar{f}_5 \alpha^3 \right) P\left( -\alpha^2 \vp^1, (\alpha^2-1)\vp^2\right) = 0 \;, \label{genphisol}
\ea
which at the GLG point reduces to 
\be
\left(\bar{f}_0 - \bar{f}_4 - \bar{f}_6 \right) + \left( \bar{f}_1 - \bar{f}_5 - \bar{f}_7 \right) \alpha + \left(\bar{f}_2 + \bar{f}_4 \right)\alpha^2 + \left( \bar{f}_3 + \bar{f}_5 \right) \alpha^3  = 0 \;.
\ee
This is solved by taking 
\be
\bar{f}_0 = \bar{f}_4 + \bar{f}_6 \;,\;\; \bar{f}_1 = \bar{f}_5 + \bar{f}_7 \;,\;\; \bar{f}_2=-\bar{f}_4 \;,\;\; \bar{f}_3=-\bar{f}_5 \;, \label{3parafluxes}
\ee
leaving four free fluxes. The same equations hold for the fluxes $\bar{h}$ and $\bar{e}$.  

Examining (\ref{genphisol}), we see that for $\bar{f}_4, \bar{f}_5 \neq 0$ this solution only holds at the GLG point.
It can be checked that this solution indeed corresponds to the R-symmetry in (\ref{glgr})
\be
W\left( \alpha^{16}\vp^1, \alpha^{32} \vp^2 \right) = \bar{\alpha}^4 W \left( \vp^1, \vp^2 \right) \;,
\ee
which has $\vp^a=0$ as its fixed point. The case $\bar{f}_4=\bar{f}_5=0$ corresponds to the faithful eigenvalues which are periodic.
This gives the R symmetry 
\be
W\left( \vp^1, \vp^2 \right) = -W \left( \vp^1, \vp^2 \right) \;,
\ee
which has all values of $\vp^1$ and $\vp^2$ as its fixed locus. Note that the first R symmetry is statistically favoured. 

Now we turn to the F-terms for the complex-structure moduli (\ref{fbuphi}). We have
\be
\bar{f} \partial_{\vp^1} P\left( \vp^a=0 \right) =  \left(\bar{f}_0 + \bar{f}_2 - \bar{f}_6 \right) + \left( -\bar{f}_1 + \bar{f}_5 + \bar{f}_7 \right) \alpha + \left(-\bar{f}_2 - \bar{f}_4 \right)\alpha^2 + \left( \bar{f}_1 + \bar{f}_3  - \bar{f}_7 \right) \alpha^3 \;,
\ee
which, as expected, vanishes for the flux choices (\ref{3parafluxes}). We also have 
\be
\bar{f} \partial_{\vp^2} P = 3 \bar{f}_4 - 3 \bar{f}_5 \alpha^3   \;,
\ee
which does not vanish but rather fixes $T$. Note that the choice of fluxes so that it is satisfied independently of $T$, $\bar{f}_4=\bar{f}_5=0$ implies that the $\vp^a$ become flat directions. For general fluxes the solution has all the moduli fixed with $\psi=\vp^1=\vp^2=0$ and 
\ba
\phi &=& \frac{\bar{e}_4\left(\bar{f}_5+\bar{f}_6 \right) + \alpha\bar{e}_5\bar{f}_7 + \alpha^2 \bar{e}_4\bar{f}_7 - \alpha^3 \bar{e}_5 \left( \bar{f}_4 + \bar{f}_6 + \bar{f}_7 \right) - e \leftrightarrow f }{\bar{e}_4\left(\bar{h}_5+\bar{h}_6 \right) + \alpha\bar{e}_5\bar{h}_7 + \alpha^2 \bar{e}_4\bar{h}_7 - \alpha^3 \bar{e}_5 \left( \bar{h}_4 + \bar{h}_6 + \bar{h}_7 \right) - e \leftrightarrow h}\;, \\
T &=& \frac{\bar{f}_4\left(\bar{h}_5+\bar{h}_6 \right) + \alpha\bar{f}_5\bar{h}_7 + \alpha^2 \bar{f}_4\bar{h}_7 - \alpha^3 \bar{f}_5 \left( \bar{h}_4 + \bar{h}_6 + \bar{h}_7 \right) - f \leftrightarrow h }{\bar{e}_4\left(\bar{h}_5+\bar{h}_6 \right) + \alpha\bar{e}_5\bar{h}_7 + \alpha^2 \bar{e}_4\bar{h}_7 - \alpha^3 \bar{e}_5 \left( \bar{h}_4 + \bar{h}_6 + \bar{h}_7 \right) - e \leftrightarrow h}\;.
\ea
Unfortunately, since the matrix $m$ relating the Picard-Fuchs basis to the symplectic basis has not been calculated in this model it is not possible to impose the tadpole relations. However since we have $10$ free flux parameters (where we have set $e_0=h_0=0$) many solutions exist to the three tadpole equations. Since we are fixed at small values 
for the complex-structure it is not clear if we retain parametric control over the Kahler modulus and the dilaton. 

%%%%%%%%%%%%%%%%%%%%%%%%%%%%%%%%%%%%%%%%%%%%%%%%%%%%%%%%%%%%%%%%%%%%%%%%%%%%%%%%%%%%%%%%%%%%%%%%%%%%%%%%%
\subsection{Low energy supersymmetry in the non-geometric landscape}
\label{nongeoland}
%%%%%%%%%%%%%%%%%%%%%%%%%%%%%%%%%%%%%%%%%%%%%%%%%%%%%%%%%%%%%%%%%%%%%%%%%%%%%%%%%%%%%%%%%%%%%%%%%%%%%%%%%%

In this section we present some extremely basic vacuum counting. Of course we have to assume that the non-geometric vacua we have been exploring are indeed legitimate 
string vacua. The aim is to show how the arguments for low energy supersymmetry being rare in the landscape 
of type IIB CY compactifications with fluxes are modified in these scenarios. The key problem in counting vacua within the non-geometric scenarios is that the fluxes 
can have `flat directions' in flux space. These are the fluxes that are unconstrained by the tadpoles and can be taken arbitrarily large. This is in contrast with IIB CY 
compactifications where the values of the fluxes are capped by the number of orientifolds. Then such flat directions mean that the distribution is dominated by large values for that 
particular flux \cite{DeWolfe:2005uu} which does not amend itself to statistical analysis in a general way. 
For comparison however we can consider the number of vacua within a hyperboloid cutoff in flux space as in 
the geometric case \cite{Ashok:2003gk,Denef:2004ze}. We take the same volume scaling so that 
the total number of vacua $N_{\mathrm{vac}}$ within the cutoff for large $L$ is given by 
\be
N_{\mathrm{vac}} \sim \left(\sqrt{L}\right)^{M_{\mathrm{fluxes}}} \;,
\ee
where $M_{\mathrm{fluxes}}$ is the number of fluxes.

The question we want to study is, within a class of compactifications, how common are vacua preserving low energy supersymmetry through $W$ vanishing perturbatively? 
This question was addressed for the IIB CY geometric landscape in a number of papers including \cite{Dine:2004is,DeWolfe:2004ns,DeWolfe:2005gy,Dine:2005gz}. 
For the generic multiple complex-structure moduli case we can follow 
the discussion of \cite{DeWolfe:2005gy}. As discussed in section \ref{multigeo}, the CY must first satisfy the geometric conditions that the monodromy matrix should have 
unfaithful eigenvalues and that at least $\phi(d)$ should be smaller than $M_{\mathrm{fluxes}}=2b_3$. For the list of CYs in \cite{DeWolfe:2005gy}, this does not seem a particularly 
strong constraint and decreases the number of vacua only by $\O\left(1\right)$ factors. Once we restrict to these CYs we have to impose an R-symmetry. This can be of the 
form valid for all values of $\vp^a$ as in section \ref{dirsol}, but could also be of the form only valid at the GLG point as in section \ref{specsol}. In both cases the 
number of constraints on the fluxes is given by the dimension of the vector space spanned by the periods $\mathrm{dim\;}\Vol$. Hence the suppression of $W=0$ vacua from 
the total number of vacua is given by 
\be
\frac{N_{W=0}}{N_{\mathrm{vac}}} \sim L^{-\mathrm{dim\;}\Vol} \;.
\ee
The important difference between the two types of R-symmetries is that the GLG $\vp^a=0$ R-symmetry is generally much more common than the R-symmetry for all values of $\vp^a$. 
Indeed for the former we always have $\mathrm{dim\;}\Vol = \phi(d)$ whilst for the latter $\phi(d) \leq \mathrm{dim\;}\Vol \leq d$. It was argued in \cite{Dine:2005gz} that 
generically for the R-symmetry valid for all $\vp^a$, $ \frac23 d \leq \mathrm{dim\;}\Vol$. We have seen this in the explicit example of section \ref{specsol} where 
this R-symmetry required setting $\bar{f}_2=\bar{f}_3=\bar{f}_4=\bar{f}_5=0$ on top of the two constraints on $\bar{f}_0, \bar{f}_1, \bar{f}_6$ and $\bar{f}_7$ thereby 
leaving only two out of eight fluxes free. Whilst for the GLG R-symmetry we had only $\phi(12)=4$ constraints leaving four out of the eight fluxes free. This difference 
gets enhanced as the number of moduli and thereby $d$ grows. Indeed $\phi(d)$ is only constrained from below by $\sqrt{d}$ and so in those cases the enhancement in the 
number of vacua can be large. We should note however that the GLG R-symmetries have an the added requirements of unfaithful eigenvalues that do not satisfy 
the periodicity condition and do satisfy the condition (\ref{fbphivanish}). 

We now turn to the non-geometric landscape. There are a number of key differences from the geometric landscape. The first is that the total number of vacua is much larger
\be
N_{\mathrm{vac}} \sim L^{\half\left[ 4\left( h^{(2,1)}+1\right)+2h^{(1,1)}\left( h^{(2,1)}+1\right) \right]} \;.
\ee
Another important difference is that solutions, at least in the large complex-structure large volume limit, are parametrically controlled \cite{Micu:2007rd}. However 
we do not take this into account in our analysis. 
The differences we will explore is that the Kahler superfields appear in the F-terms for the complex-structure moduli (\ref{fbuphi}) and that we 
have to also solve the F-terms for the Kahler moduli (\ref{fbe}). For the purposes of this analysis 
we assume that all the Kahler moduli appear perturbatively in the superpotential, i.e. we keep the $e^k_{\Lambda}$ fluxes on for all the values of $k$. 

The two types of vacua in sections \ref{dirsol} and \ref{specsol} obey different statistics. Consider first the R-symmetries vacua 
of section \ref{specsol}. 
We have already discussed its statistics in the geometric case. The difference here is that the conditions (\ref{fbupsi}) are no longer relevant since we can 
solve the complex-structure F-terms using the Kahler moduli. However we now have the $h^{(1,1)}$ equations (\ref{fbe}) to solve. Therefore the suppression factor is
\be
\frac{N_{W=0}}{N_{\mathrm{vac}}} \sim L^{-\half\left( h^{(1,1)}+2\right)\phi(d)} \;.
\ee
This behaves in very much the same way as the geometric suppression where although the suppression grows with the number of moduli, it does so slower than 
the total number of vacua so that for large numbers of moduli the number of $W=0$ vacua scale like the total number of vacua. 

We now turn to the direct solution vacua of section \ref{dirsol}. In doing this we assume that the CY has some point in its moduli space where the dimension of 
$\Vol$ reduces to some $\mathrm{dim\;}\Vol \equiv \D < b^3$. Then the number of vacua is suppressed as 
\be
\frac{N_{W=0}}{N_{\mathrm{vac}}} \sim L^{-\half\left( h^{(1,1)}-h^{(2,1)}+3\right)\D} \;.
\ee
This follows since the $h^{(1,1)}+2$ equations (\ref{fbh}), (\ref{fbf}) and (\ref{fbuphi}) are in $h^{(2,1)}-1$ variables which implies that $h^{(1,1)}-h^{(2,1)}+3$ of them must be solved by picking fluxes. Recall that a condition for these vacua to exist is $h^{(1,1)} \ge h^{(2,1)} - 1$. It is not very clear what the value for $\D$ should be for generic numbers of moduli. However it is worth noting that the suppression factor for these vacua can be relatively small for $h^{(1,1)}\sim h^{(2,1)}$.

To summarise, within non-geometric compactifications $W=0$ vacua are suppressed compared to the overall number of vacua but not as drastically as in the geometric case.  

%%%%%%%%%%%%%%%%%%%%%%%%%%%%%%%%%%%%%%%%%%%%%%%%%%%%%%%%%%%%%%%%%%%%%%%%%%%%%%%%%%%%%%%%%%%%%%%%%%%%%%%%%
\section{Non-perturbative effects in non-geometric compactifications}
\label{nonper}
%%%%%%%%%%%%%%%%%%%%%%%%%%%%%%%%%%%%%%%%%%%%%%%%%%%%%%%%%%%%%%%%%%%%%%%%%%%%%%%%%%%%%%%%%%%%%%%%%%%%%%%%%%

In this section we consider non-perturbative effects in the non-geometric compactifications we have been studying. 
These effects correspond to branes wrapping cycles which in non-geometric settings are still to be fully understood. The formalism to describe these branes is under
development \cite{Evslin:2007ti,Koerber:2007xk}. 
Roughly speaking, in a non-geometric, or generalised geometric setting, we can think of generalised cycles as being composed of chains of cycles 
of even or odd degrees. Then networks of branes of even or odd dimensions can wrap these cycles. Effectively, however, in terms of the superpotential, we can 
simply think of the effects as being induced by the usual geometric branes. 
The particular effects are E3 instantons and gaugino condensation on D7 branes in IIB, which induce terms in the superpotential that are exponential in the Kahler moduli \cite{Becker:1995kb,Witten:1996bn,Harvey:1999as}. 
Their mirrors are E2 instantons and gaugino condensation 
on D6 branes in IIA. The aim of this section is to study, in parallel, the constraints on the presence of 
such effects imposed by supergravity gaugings and by world-volume anomalies. We show that these constraints are 
compatible with the presence of non-perturbative effects in the same moduli that appear perturbatively. We then 
show that such effects perturb the supersymmetric Minkowski vacua found in section \ref{susymink} to supersymmetric AdS vacua but with an exponentially small gravitino mass. 
This is the hierarchy generating mechanism that we argued for in the introduction. 

We also include a discussion regarding how implementing an intersecting branes sector may not be compatible with the scenarios in this paper where all the 
moduli are fixed perturbatively. This arises if closed string axions are needed to cancel anomoulus $U(1)$s. 
We discuss the flux choices that are needed for the presence of axions in the theory and the resulting effective KKLT-like theories. 

%%%%%%%%%%%%%%%%%%%%%%%%%%%%%%%%%%%%%%%%%%%%%%%%%%%%%%%%%%%%%%%%%%%%%%%%%%%%%%%%%%%%%%%%%%%%%%%%%%%%%%%%%
\subsection{Supergravity gauging constraints}
\label{sugcon}
%%%%%%%%%%%%%%%%%%%%%%%%%%%%%%%%%%%%%%%%%%%%%%%%%%%%%%%%%%%%%%%%%%%%%%%%%%%%%%%%%%%%%%%%%%%%%%%%%%%%%%%%%%

It is simpler to study these effects in the type IIA set-up. We return to the IIB case later. 
We are interested in effects that arise from E2/D6 branes wrapping three-cycles so that they induce a term in the superpotential of type
\be
W^{IIA} \sim \sum_{k} A_k\left( T^a,S \right) e^{ia^k U_k} \;, \label{iianonper}
\ee
where $A_k$ are some holomorphic functions and the $a^k$ are real positive constants. 
Recall that before the orientifold truncation the four-dimensional theory is an $\N=2$ supergravity. The only way to induce a potential in these supergravities is to gauge isometries in the manifold formed by the scalar field values. Since the $\N=1$ superpotential arises as a truncation of the $\N=2$ theory it is 
useful to keep this origin in mind. Let us consider an example where we have turned off the metric, non-geometric, and RR fluxes. There is a term in the superpotential given by (\ref{iiansw}) which we reproduce here for convenience 
\be
W_{IIA}^{'} = -e_{h^{(2,1)}+1}^k U_k \label{iiansw2}\;.
\ee
The origin of this term can be traced back to gauging the shift isometry of one of the $\N=2$ scalar superpartners of the $\nu_k=\re{U_k}$ axions which we denote $\tilde{\nu}_k$. These fields come from the expansion of the RR three-form as 
\be
C_3 = \nu_k \beta^k + \tilde{\nu}^k \alpha_k + \nu^0 \alpha_0 + \tilde{\nu}_0 \beta^0 \;.
\ee
Here $\alpha_0$, $\alpha_k$, $\beta^0$ and $\beta^k$ denote a basis of three-forms. 
The forms $\alpha_k$ and $\beta^0$ are odd under the orientifold action and the $\alpha_0$, $\beta^k$ are even. Since $C_3$ is even, the fields $\tilde{\nu}^k$ are projected out and do not appear in the $\N=1$ theory. In the $\N=2$ theory they become gauged as \cite{Louis:2002ny}
\be
D_{\tilde{\nu}^k} = \partial \tilde{\nu}^k - e_{h^{(2,1)}+1}^k V^{h^{(2,1)}+1} \;,
\ee
where $V^{h^{(2,1)}+1}$ is the $\N=2$ graviphoton and the NS flux is 
\be
H = -e_{h^{(2,1)}+1}^k \alpha_k \;. \label{nsiiaflux} 
\ee 
If non perturbative effects of the form $e^{i\tilde{\nu}^k}$ were present they would break the gauged shift 
isometry $\tilde{\nu}^k \ra \tilde{\nu}^k + a^k$. 
Therefore supergravity constraints rule out such effects in the presence of the 
flux $e_{h^{(2,1)}+1}^k$ \footnote{This constraint makes for an interesting no-go theorem for 
supersymmetric Minkowski vacua from type II compactifications without orientifolds with all the geometric moduli stabilised. This follows since to introduce a potential 
we must gauge an isometry which implies that at least one axion must be massless and therefore its geometric supersymmetric partner must also be massless.}. As we have seen this constraint is automatically satisfied by the orientifold since the fields $\tilde{\nu}^k$ are projected out.

As well as protecting the shift isometry of $\tilde{\nu}^k$, the flux $e_{h^{(2,1)}+1}^k$ also breaks the shift isometries of the $\nu_k$. However this does not mean that they are protected against non-perturbative effects. 
In this case these would be E2-branes wrapping the homological dual of $\beta^k$. We see that the orientifold 
makes sure that the fluxes present no constraints on the presence of non-perturbative effects in the fields that 
are left by the orientifold projection.

Consider turning on the NS $h_{h^{(2,1)}+1}$ and metric $h_{a}$, $e_{a}^k$ fluxes. These can also be traced to gauging isometries in the $\N=2$ theory. They pair up with $e_{h^{(2,1)}+1}^k$ to form a matrix $e_A^K$, where $K=0,k$, which appears in the 
gauge derivative as
\be
D_{\tilde{\nu}^K} = \partial \tilde{\nu}^K - e_{A}^K V^A \;, \label{sugraind}
\ee
where $V^A$ are composed of the graviphoton and the vector in the $h^{(1,1)}$ vector-multiplets $V^a$. We see that they 
gauge the same isometries but just with respect to all the gauge fields. So that they do not impose any further constraints on the presence of non-perturbative effects.

We also have the non-geometric fluxes $h_{\tilde{a}}$ and $e_{\ta}^k$. It turns out that these also gauge the same 
isometries \cite{D'Auria:2007ay}. They act as magnetic charges for the fields $\tilde{\nu}^K$ just as the metric and NS fluxes acted as electric charges. So in their presence the fields $\tilde{\nu}^K$ are dyons. To see this we have to dualise the 
fields $\tilde{\nu}^K$ into two-forms $\tilde{B}_K$. This is a procedure developed in \cite{Dall'Agata:2003yr,D'Auria:2004yi} for writing supergravities with dyons. Recall that in four-dimensions a massless scalar is dual to a massless two-form. Then the non-geometric fluxes act as masses for the two-forms, which we turn on after dualising. 
These appear by modifying the gauge field-strengths $F^A$ for the $V^A$ as 
\be
F^A \rightarrow F^A + e_{\tilde{A}}^K B_K \;.
\ee
The key point is that this procedure can only be performed if we have the shift isometries in the $\tilde{\nu}^K$ and so these isometries are protected by these fluxes. 

To summarise, supergravity gauging imposes constraints on the presence of non-perturbative effects in general, but we have shown here that the orientifold automatically makes sure that the isometries gauged by all the fluxes present (including non-geometric fluxes) are precisely in the fields that are projected out. This guarantees that after the 
orientifold projection there are no further constraints from supergravity gaugings on the presence of non-perturbative 
effects. 

We now turn to explaining the same results in terms of world-volume anomalies on the branes that give rise to the non-perturbative effects.

%%%%%%%%%%%%%%%%%%%%%%%%%%%%%%%%%%%%%%%%%%%%%%%%%%%%%%%%%%%%%%%%%%%%%%%%%%%%%%%%%%%%%%%%%%%%%%%%%%%%%%%%%
\subsection{World-volume anomalies}
%%%%%%%%%%%%%%%%%%%%%%%%%%%%%%%%%%%%%%%%%%%%%%%%%%%%%%%%%%%%%%%%%%%%%%%%%%%%%%%%%%%%%%%%%%%%%%%%%%%%%%%%%%

It was shown in \cite{Kashani-Poor:2005si}, using world-volume anomalies, that isometries that are gauged by NS fluxes are protected against non-perturbative effects that may break them. We now show that this applies to all the fluxes. For important related discussions which we return to in section \ref{chirmat} see \cite{Villadoro:2006ia,Haack:2006cy}. 

Consider first the presence of just NS flux (\ref{nsiiaflux}). Then we have argued that non-perturbative effects that 
would give rise to a term $e^{i\tilde{\nu}^k}$, which correspond to branes wrapping the cycle $\alpha_k$, are forbidden.
Recall that if a D-brane wraps a cycle with H-flux through it, the D-brane gauge field, with field strength $F^D$, satisfies the Bianchi Identity
\be
dF^D + H^{NS} = 0 \;.
\ee
This violates Gauss' law on the world-volume of the D-brane. We henceforth refer to such inconsistency as the Freed-Witten (FW) anomaly \cite{Freed:1999vc}.  Therefore we recover that branes can not wrap $\alpha_k$ since these are precisely 
the cycles through which there is NS flux. 

Now consider turning on the full NS and metric fluxes $e_A^k$. The metric fluxes appear as parameters of the torsion of the manifold and measure the failure of the basis 
forms to be harmonic
\ba
d\alpha_{0} &=&  - h_{a} \tilde{\omega}^a \;,\\ \nn
d\beta^{k} &=&  - e_a^{\;\;k} \tilde{\omega}^a \;,\\ \nn
d\omega_a &=& - h_{a} \beta^0 + e_a^{\;\;k} \alpha_k \;.
\ea
We argued that these again protect us against the same non-perturbative effects. In the presence of these fluxes the world-volume Gauss' law is modified to 
\be
dF^D + H^{NS} - dJ_c = 0 \;, \label{fwmetric}
\ee 
where $J_c=-B+iJ=T^a \omega_a$. We can interpret the new term as the condition that the brane should wrap a sub-manifold without a boundary, i.e. a cycle \cite{Villadoro:2006ia}. 
The metric fluxes mean that the cycles $\beta^0$ and $\alpha_k$ have a boundary and so rule out the non-perturbative effects in 
$\tilde{\nu}^k$ because they would correspond to branes wrapping sub-manifolds with boundaries. 

We now consider non-geometric fluxes $h_{\ta}$ and $e_{\ta}^k$. To understand how they feature we introduce some 
notation that matches the compactifications studied in \cite{Micu:2007rd}. 
The non-geometric fluxes arise from the action of a modified derivative operator $\D$, that can be thought of as covariant derivative for T-dualities and also includes the NS flux, as 
\ba
\D\alpha_{0} &=& h_{\ta} \omega_a - h_{a} \tilde{\omega}^a + h_{h^{(2,1)}+1} \epsilon \;, \nn \\
\D\beta^{k} &=& e_{\ta}^k \omega_a - e_a^{\;\;k} \tilde{\omega}^a  + e_{h^{(2,1)}+1}^k \epsilon \;, \nn \\ 
\D\omega_a &=& - h_{a} \beta^0 + e_a^{\;\;k} \alpha_k \;, \nn \\
\D\tilde{\omega}^a &=& -h_{\ta} \beta^{0} + e_{\ta}^{\;k} \alpha_k \;. \label{diffiia}
\ea
In terms of the derivative operator we can propose a neat extension to (\ref{fwmetric}) 
\be
dF^D - \D \Pi^{ev}= 0 \;. \label{fwnongeo}
\ee
Here $\Pi^{ev}=e^{J_c}$ which is a combination of zero, two, four and six-forms. It can be thought of as a pure spinor\footnote{Recall that a spinor is equivalent to a set of forms of different degrees that can be generated by acting on it with anti-symmetric combinations of gamma matrices.} associated with $SU(3) \times SU(3)$ compactifications, 
see \cite{Grana:2006kf} for a nice review.
In this way we can interpret the condition as requiring the brane to be wrapped on a non-geometric 'cycle'. 
In evaluating (\ref{fwnongeo}) we should pick out the three-form component of $\D \Pi^{ev}$.
It is simple to check that this is only proportional to $\beta^0$ and $\alpha_k$. Therefore the full fluxes 
only constrain branes wrapped on those cycles and so do not constrain non-perturbative effects in $\nu^k$. 
This matches the supergravity gauging constraints and we can now clearly see why the orientifold guarantees this. 
The quantity $\D \Pi^{ev}$ has a definite parity under the orientifold action which is the opposite one to the spinor 
that gives rise to the superfields
\be
\Pi_c^{od} = S \alpha_0 + U_k \beta^k \;. \label{picod}
\ee

The formulation (\ref{fwnongeo}) can easily be applied to IIB now since mirror symmetry simply interchanges the spinors 
$\Pi$ for some other spinors $\Psi$ as outlined in \cite{Grana:2005ny}. Again the orientifold parity implies that the fluxes impose no constraints on the presence of non-perturbative effects which are E3 branes and D7 branes wrapped on four-cycles.

To summarise we argued that we expect non-perturbative effects, in the same fields that appear perturbatively, to be 
present in the compactifications. 

%%%%%%%%%%%%%%%%%%%%%%%%%%%%%%%%%%%%%%%%%%%%%%%%%%%%%%%%%%%%%%%%%%%%%%%%%%%%%%%%%%%%%%%%%%%%%%%%%%%%%%%%%
\subsection{The non-perturbative vacuum}
\label{nonpervac}
%%%%%%%%%%%%%%%%%%%%%%%%%%%%%%%%%%%%%%%%%%%%%%%%%%%%%%%%%%%%%%%%%%%%%%%%%%%%%%%%%%%%%%%%%%%%%%%%%%%%%%%%%%

We now consider the effect that the non-perturbative corrections to the superpotential have on the supersymmetric Minkowski vacua of the perturbative superpotential. 
We assume the perturbative vacua were at large, and perhaps parametrically controlled, values for the fields which means we can consider the non-perturbative effect as a small perturbation of the vacuum. 
The relevant perturbative parameter is the value of the non-perturbative part of the superpotential evaluated at the values of the 
superfields in the perturbative vacuum. Let us denote the superfields collectively as $\phi_i$, and their values in the original perturbative Minkowski vacuum as $\phi^0_i$ 
where the index $i$ ranges over the type of superfield. The perturbative parameter is $\left. W^{NP} \right|_{\phi^0_i} = Ae^{ia\phi^0_{U}} \equiv r$. For a TeV scale gravitino we require $r \sim 10^{-13}$. Such a perturbation can not destabilise the perturbatively stable vacuum (which had mass matrix eigenvalues of order one) and so there should still be a stable vacuum but at slightly shifted values of the fields. To see the degree by which the fields shift we write the scalar potential as
\be
V = V^P + V^{NP} \;,
\ee
where $V^{NP}$ is a contribution which is induced by the non-perturbative part of the superpotential, and $V^P$ is the original perturbative scalar 
potential. Now we consider expanding $\partial_{\phi_i}V$ about the perturbative vacuum 
\be
\partial_{\phi_i}V = \left. \partial_{\phi_i}V^P \right|_{\phi_m^0} + \left. \partial_{\phi_j}\partial_{\phi_i}V^P \right|_{\phi_m^0} \delta \phi_j 
 + \left. \partial_{\phi_i}V^{NP} \right|_{\phi_m^0} +  \left. \partial_{\phi_j}\partial_{\phi_i}V^{NP} \right|_{\phi_m^0} \delta \phi_j + \O\left( \delta \phi^2 \right) \;. 
\ee
In the new minimum this should vanish. The first term on the right hand side vanishes. The second term is just the mass matrix for the original vacuum which, since 
we assume that the original vacuum is perturbatively stable, is non-vanishing. This implies that it must cancel against terms of order $r$ (or higher powers of $r$) in the 
non-perturbative part. Hence $\delta \phi$ must be at least as small as $r$ up to factors of $\O(1)$.  
We are interested in the value of the superpotential in the new vacuum $W_0$ and this is given by 
\ba
W_0 &=& \left. W_{P} \right|_{\phi^0_i} + \left. W_{NP} \right|_{\phi^0_i} + \left. \partial W_{P} \right|_{\phi^0_i} \O\left(r\right) + \left. \partial W_{NP} \right|_{\phi^0_i} \O\left( r \right) + \O \left( r^2 \right) \nn \\
&=& \left. W_{NP} \right|_{\phi^0_i} + \O \left( r^2 \right) \nn \\ 
&\cong & r\;,
\ea
where we used the fact that in the original supersymmetric Minkowksi vacuum the superpotential and its derivatives vanished.
As expected, given the original perturbatively stable supersymmetric Minkowski vacuum,  
we have recovered a vacuum where all the moduli are stabilised and where the value of the gravitino mass is dynamically small.

We have yet to determine whether supersymmetry is preserved or broken in the perturbed vacuum. Since the value of the superpotential is non-vanishing we certainly 
no longer have a supersymmetric Minkowski vacuum. To determine further properties of the vacuum explicitely requires knowledge of the form of the Kahler potential. 
This is likely to be a complicated function once warping effects and perturabtive and non-perturbative corrections, which should be large for these 
vacua where the manifold is non-geometric and the fields lie close to the LG point, are included. Further such an analysis would also require knowledge of the function $A$ 
which multiplies the non-perturbative effects since, unlike in KKLT, this can not be integrated out. Therefore actually solving for the F-terms is a task beyond the 
scope of this paper. Generically however we would expect that supersymmetry is preserved 
in the new vacuum in which case it is AdS. The reason is that the F-terms now include all the fields and so there are as many equations to solve as there are fields 
(non-superysmmetric vacua are usually associated with overconstrained sets of equations). 
Also since the fields are only slightly perturbed the new values are still physical.

We should note that the vacuum analysis performed in this section is based on exploring small perturbations of the original sueprsymmetric Minkowski vacuum and 
so is not applicable to any vacua, supersymmetric or not, which are far away from the original vacuum. 

%%%%%%%%%%%%%%%%%%%%%%%%%%%%%%%%%%%%%%%%%%%%%%%%%%%%%%%%%%%%%%%%%%%%%%%%%%%%%%%%%%%%%%%%%%%%%%%%%%%%%%%%%
\subsection{Chiral matter}
\label{chirmat}
%%%%%%%%%%%%%%%%%%%%%%%%%%%%%%%%%%%%%%%%%%%%%%%%%%%%%%%%%%%%%%%%%%%%%%%%%%%%%%%%%%%%%%%%%%%%%%%%%%%%%%%%%%

The type of constructions outlined so far in this paper have a low gravitino mass and, if the constraints on the number of moduli in section \ref{dirsol} and \ref{specsol} are saturated, all the moduli fixed at the string scale. However 
eventually we would like to introduce an open string sector in which we hope 
to realise the standard model. This could arise from intersecting D6 or D8 branes in the type IIA case and from magnetised D7-branes or D3 branes on singularities in type IIB. We will not concern ourselves with the details of these models but only in the implications of their presence on the closed string moduli. We will see that most constructions require us to turn off some fluxes so as to destabilise at least one modulus. 

Most of the current constructions of realistic brane models contain anomalous $U(1)$ gauge fields. In the standard 
constructions their anomalies are cancelled through the Green-Schwartz (GS) mechanism. Let us outline how this works 
Consider turning off all fluxes and wrapping a D6 brane on the cycle, $A_k$, dual to the form $\alpha_k$.
Now let us also include some other D6 branes that intersect this D6 brane at angles. The chiral gauge theory living on the intersection has an anomalous $U(1)$. However, there is a term on the D6 world-volume which induces a GS coupling
\be
S_{D6} \supset \int_{M_4 \times A_k}{C_5 \wedge F^D} = \int_{M_4}{C^k_2 \w F^D} \;,
\ee
where $C_5$ is the RR five-form which we expand as $C_5 = C^k_2 \alpha_k$ and $F^D$ is the field strength of the world-volume gauge field. 
The four-dimensional tensors $C^k_2$ are dual to the RR axions $\nu_k$. Then the GS mechanism means that the anomalous  $U(1)$ eats the axion and becomes massive, turning into a global symmetry in the low energy effective theory. 

The essential problem is that this requires the $\nu_k$ to be massless while generically they will be massive from the 
fluxes. This implies we have to turn off some fluxes which we study in section \ref{axflux}.

Before proceeding it is worth mentioning that requiring axions from the closed string sector is a model dependent effect. It is possible that the axions 
could come from a different sector such as twisted states as occurs for D3 branes on singularities \cite{Douglas:1996sw}. Also brane networks, which are natural in 
non-geometry could be promising future avenues \cite{Evslin:2007ti}. Therefore, although we present a brief outline 
of the consequences of such axions in the closed string sector we should keep in mind that their presence is not certain.

%%%%%%%%%%%%%%%%%%%%%%%%%%%%%%%%%%%%%%%%%%%%%%%%%%%%%%%%%%%%%%%%%%%%%%%%%%%%%%%%%%%%%%%%%%%%%%%%%%%%%%%%%
\subsubsection{Axions and fluxes}
\label{axflux}
%%%%%%%%%%%%%%%%%%%%%%%%%%%%%%%%%%%%%%%%%%%%%%%%%%%%%%%%%%%%%%%%%%%%%%%%%%%%%%%%%%%%%%%%%%%%%%%%%%%%%%%%%%

In this section we study the constraints on the fluxes imposed by requiring flat directions to cancel the $U(1)$ anomalies. 
The axion shift symmetry could be a full symmetry of the action or it could be an enhanced symmetry 
of the vacuum. This distinction is generally important in determining the constraints on the fluxes. The vacuum 
symmetry is less strict than the full symmetry in terms of having to switch off fluxes. In particular it means that 
in AdS it is possible to fix the geometric moduli and leave flat axionic directions in the vacuum \cite{Camara:2005dc}. It is possible that this situation could also arise in a non-supersymmetric vacuum. However, it was argued in \cite{Villadoro:2006ia} that the gauge invariance should be at the action level. For our purposes there is little 
practical difference since we are studying supersymmetric Minkowski vacua which means that any flat axionic directions in the vacuum imply a massless geometric superpartner. This means that with regards to finding fully stabilised vacua we might as well impose the axionic symmetry as a symmetry of the action. This is the approach we take in this section.

In IIA the candidate axion fields are the real part of the superfields arising from (\ref{picod}). In 
the purely geometric case, with no H-flux, the constraint on them being axions is that the forms they are 
expanded in are closed. Let us denote a linear combination, labelled by $i$, of $\re{S}$ and $\re{U_k}$ by a vector $A^{(i)}_K$, where 
we use the index notation as in sections \ref{4dsuperpot} and \ref{sugcon}. The entries of $A^{(i)}_K$ correspond to which axions appear in the linear combination. Then the constraint on being closed reads 
\be
d \left( A^{(i)}_0 \alpha_0 + A^{(i)}_k \beta^k \right) = 0 \;,
\ee
which can be written as
\be
e^{\;\;K}_a A^{(i)}_K = \left(e A^{(i)}\right)_a = 0 \;.
\ee
If we also turn on the H flux and the non-geometric fluxes this generalises to 
\be
e^{\;\;K}_{\Lambda} A^{(i)}_K = \left( e A^{(i)} \right)_{\Lambda} = 0 \;. \label{unfix}
\ee
Recall that in IIA the index $\Lambda$ runs over the even forms and has dimension $2\left(h^{(1,1)}+1\right)$.
We see that, for general fluxes, out of the possible $h^{(2,1)}+1$ axions $2\left(h^{(1,1)}+1\right)$ get fixed.
Note that in a supersymmetric Minkowski vacuum, as argued in section \ref{perfer}, there is an enhanced symmetry in the 
vacuum which implies that only $h^{(1,1)}+1$ axions get fixed. If $2\left(h^{(1,1)}+1\right) > h^{(2,1)}+1$ then 
we have to turn off a row of fluxes in the matrix $e$ for each unfixed axion we require. 

To summarise, for our models, the presence of axions requires that the fluxes and number of moduli are such that 
some superfields do not appear in the perturbative superpotential.  

It is worth noting that the world-volume anomaly constraint (\ref{fwnongeo}) and the unfixed axion constraint (\ref{unfix}) are the same if we impose them at the action level, i.e. for all values of the $T^a$, but differ if we impose them at the 
vacuum level. Indeed in a Minkowski vacuum (\ref{fwnongeo}) is always satisfied but in general there are no axions. 

%%%%%%%%%%%%%%%%%%%%%%%%%%%%%%%%%%%%%%%%%%%%%%%%%%%%%%%%%%%%%%%%%%%%%%%%%%%%%%%%%%%%%%%%%%%%%%%%%%%%%%%%%
\subsubsection{Generalised KKLT}
\label{genKKLT}
%%%%%%%%%%%%%%%%%%%%%%%%%%%%%%%%%%%%%%%%%%%%%%%%%%%%%%%%%%%%%%%%%%%%%%%%%%%%%%%%%%%%%%%%%%%%%%%%%%%%%%%%%%

The problem discussed above implies that it could be that in order to include chiral D6 branes we should have another sector to the theory with perturbatively massless superfields\footnote{Note that forcing the corresponding fluxes to vanish so that these fields to not appear perturbatively in the superpotential is very costly in terms of the landscape. If constructions were found where standard model like sectors could be realised without needing axions to cancel anomalies they would be statistically favoured.}. The real parts of these superfields will now be proper axions with shift symmetries and these can be used to cancel the anomalies on the brane world-volume. 
Although the new superfields do not appear in the superpotential perturbatively they can appear non-perturbatively through E2/D6 brane effects\footnote{Non-perturbative effects break the shift symmetries of the axions to integers. However, in the presence of a chiral open string sector the open string fields appear in the superpotential in such a way to restore the full gauge invariance under shifts so that the actual axions that participates in the 
GS mechanism is a combination of the closed string axion and the open string fields \cite{Villadoro:2006ia,Haack:2006cy,Achucarro:2006zf,Cremades:2007ig}.}. 

The resulting closed string sector resembles KKLT constructions in the sense that we have a sector of moduli 
that are fixed perturbatively and a sector that is fixed non-perturbatively. 
 Although we have discussed these constraints in the type IIA framework, they apply equally to type IIB setups where again we require axions to cancel anomalies on magnetised D7 branes. 
 Therefore, for concreteness, we henceforth use type IIB notation. 
Combining the superpotentials of the previous two sections, the type of setups that arise are schematically of the form
\be
W^{IIB}\left(S,U,T_1,T_2\right) = W^P\left(S,U,T_2\right) + B\left(S,U\right)e^{ibT_2} + A\left(S,U\right)e^{iaT_1}  \;. \label{waxionschem}
\ee
The real part of the field $T_1$ is an axion that can participate in the anomaly cancellation. The key point is that it does not appear in the perturbative superpotential $W^P\left(S,U,T_2\right)$. We are interested in the vacua where $T_2$ is stabilised perturbatively at some large value in a $W^P=0$ vacuum. Then, as discussed in the previous section, introducing the second term in (\ref{waxionschem}) leads to an AdS vacuum with $W \sim Be^{ibT_2} \equiv \delta$. 
Since the fields $S, U$ and $T_2$ all appear perturbatively in the superpotential their masses will be of order the string scale. 
While the field $T_1$ only appears non-perturbatively and so we expect it to be nearer the gravitino mass. 
Indeed, by diagonalising the mass matrix it is possible to show that the mass of $T_1$ is of order $\mathrm{ln\;}\left(M_p/m_{3/2}\right)m_{3/2}$ up to order one 
factors induced by mixing with the superfield $T_2$. If we integrate out the string scale fields, by 
replacing them with their values as given by the perturbative part of the superpotential,  
we are left with an effective KKLT-like superpotential 
\be
W^{IIB} = W_0 + Ae^{iaT_1}  \;, \label{waxionschem2}
\ee
where $W_0 \sim \delta$. Then $T_1$ will be fixed such that $Ae^{iaT_1} \sim \delta$ and so its mass will be much lower than the string scale for a hierarchically small $\delta$ making the approximation of the effective theory valid. 

There are some differences between the above constructions and the original KKLT ones. 
The moduli that are stabilised non-perturbatively 
are not decoupled in the Kahler potential from the perturbatively fixed ones. In the KKLT scenario, up to loop effects \cite{Berg:2005ja}, the complex-structure moduli are decoupled from the Kahler moduli in the Kahler potential and this effect is absent. In these scenarios integrating out some of the Kahler/complex-structure moduli while leaving others leads to non-canonical effective Kahler potentials for those moduli. Such effects may be relevant for supersymmetry breaking through corrections to the Kahler potentials. 
Another difference is that in the above constructions we could have parametric control over the vacuum values for the moduli. Indeed for the toy example studied 
in section \ref{dirsol}, we could show that was the case for at least the complex-structure moduli. 
This control is important in generating small parametric hierarchies. These 
include  usual ones between the string and KK scale and the moduli masses. But may also include a mass hierarchy between the non-perturbatively fixed moduli, $T_1$ in the 
example above, and the gravitino mass. This could be useful in avoiding light moduli and their associated problems. Finally these vacua do not suffer from the KKLT  
stability issues as studied in \cite{Choi:2004sx} since the vacuum is a perturbed supersymmetric Minkowski vacuum which is stable. 
Apart from the effects discussed above, the moduli stabilisation scheme is the same as that of KKLT.

%%%%%%%%%%%%%%%%%%%%%%%%%%%%%%%%%%%%%%%%%%%%%%%%%%%%%%%%%%%%%%%%%%%%%%%%%%%%%%%%%%%%%%%%%%%%%%%%%%%%%%%%%
\section{Supersymmetry breaking}
\label{susybreak}
%%%%%%%%%%%%%%%%%%%%%%%%%%%%%%%%%%%%%%%%%%%%%%%%%%%%%%%%%%%%%%%%%%%%%%%%%%%%%%%%%%%%%%%%%%%%%%%%%%%%%%%%%%

Obtaining a dynamically low gravitino mass allows for the possibility of a low scale of supersymmetry breaking. 
So far supersymmetry has been preserved and so we need to include new effects to break supersymmetry. 
In this section we do not present new methods for this but study how some of the current suggestions can be implemented within 
the framework of this paper. A natural requirement is that these mechanisms also uplift the now AdS vacua back up to Minkowski vacua. 

The main issue in breaking supersymmetry within these models is how to keep the supersymmetry breaking effects to a dynamically low scale so that the low gravitino 
mass is not washed out. This is automatic if, as discussed in section \ref{nonpervac}, once the non-perturabtive effects are introduced, 
the new perturbed vacuum is non-supersymmetric. Such a set-up is likely to rely on corrections to the Kahler potential as in \cite{Davidse:2005ef,Dudas:2006gr}. 

Another possibility is to include, in IIB, anti D3 branes in a warped region as in the original KKLT 
scenario \cite{Kachru:2003aw}. These break supersymmetry at a scale suppressed by the warp factor at their location. Therefore 
by picking the warp factor to be of order the gravitino mass such effects do not wash out the low scale. This is 
an example of balancing out two dynamically generated scales which features in all the upcoming constructions. 

A manifestly supersymmetric version of the anti brane mechanism is D-term supersymmetry breaking induced by brane world-volume fluxes \cite{Burgess:2003ic}. Since the D-term is generated 
by one of the axions getting charged we require, as in section \ref{chirmat}, that the axion shifts remain a symmetry and 
so they should not appear perturbatively in the superpotential. However in both IIA and IIB settings they can appear 
non-perturbatively through E2 and E3 branes and through gaugino condensation on D6 and D7 branes. This allows for their real parts to be fixed along with the uplift. 
Such scenarios were studied in \cite{Cremades:2007ig} within a type IIB context, but can equally be realised in a type IIA mirror setup with D6 branes taking the role of the D7 branes. These  models of D-term uplifting can be separated into two classes. Those where the charges of the open string sector are not all the same sign, and those where they are all of the same sign. The former models, to which all the explicit examples belong, have the property that the open string fields can adjust their values to make the D-terms vanish\footnote{This can be avoided to some extent by not including one of the fields in the superpotential as in section 3.2 of \cite{Cremades:2007ig}. With this mechanism however the D-terms scale as $D^2 \sim m_{3/2}^4$ and so can not uplift the negative $m_{3/2}^2M_p^2$ cosmological constant.}. The second class of models is where all the fields have the same sign. In that case the fields can not cancel and the D-terms are of order $M_p^4$ and so they wash out the dynamically low gravitino mass minimum. However, as pointed out in \cite{Cremades:2007ig}, the D-terms can be suppressed by warping effects by placing the D-branes in a highly warped region. In that case the warping can be chosen to lower the D-term scale to that of the gravitino mass scale. Note that the warping suppression is only really well understood in the IIB.

%%%%%%%%%%%%%%%%%%%%%%%%%%%%%%%%%%%%%%%%%%%%%%%%%%%%%%%%%%%%%%%%%%%%%%%%%%%%%%%%%%%%%%%%%%%%%%%%%%%%%%%%%
\subsection*{Metastable vacua}
%%%%%%%%%%%%%%%%%%%%%%%%%%%%%%%%%%%%%%%%%%%%%%%%%%%%%%%%%%%%%%%%%%%%%%%%%%%%%%%%%%%%%%%%%%%%%%%%%%%%%%%%%%

It is also possible to break supersymmetry in a metastable vacuum of an open string sector along the lines of ISS \cite{Intriligator:2006dd}. 
Concrete realisations of an appropriate gauge theory that contains a metastable 
vacuum within compactifications of string theory have yet to be constructed. For some recent attempts see \cite{Franco:2006ht}. We can highlight the important 
features that are relevant for the non-geometric vacua we have been studying through a toy model. Consider the following theory that corresponds to adding 
an ISS sector to the closed string sector of the previous sections
\ba
K &=& K^T\left( T_1,\bar{T_1},T_2,\bar{T_2}\right) + K^S\left(S,\bar{S}\right)+ K^U\left(U,\bar{U}\right) + \left| \vp \right|^2 + \left| \tilde{\vp} \right|^2 + \left| \Phi \right|^2 \;, \nn \\
W &=& W^P\left(U,S,T_1,T_2\right) + h\mathrm{Tr\;} \tilde{\vp}\Phi\vp - hmAe^{iaT_1}\mathrm{Tr\;} \Phi + Be^{ibT_2} ;. \label{isstoy}
\ea
Here the open string sector which will be responsible for triggering supersymmetry breaking is given by an $SU\left(n\right)$ gauge theory which in the magnetic description has a description in terms of the degrees of freedom $\vp^a_i$, $\tilde{\vp}^a_i$, $\Phi^i_j$ where $a$ is a colour index $a=1,...,n$ and $i$ is a flavour index $i=1,...,N_f$. In terms of the electric theory the $\vp$ are related to Baryons and $\Phi$ to Mesons and the magnetic gauge group $SU(n)$ is related to the electric gauge group $SU(N_c)$ through $n=N_f-N_c$. The magnetic theory has a perturbative description in the free magnetic range $N_f>3n$. The strong dynamics energy scale of the magnetic theory $\Lambda$ is the same as that of the electric theory and is given by $\left|Ae^{iaT_1}\right|$. The parameter $m$ is a small $m \ll \Lambda$ mass term introduced for the quarks. For possible ideas as to the origin of such a small parameter see \cite{Dine:2006gm} where it can also be realised dynamically. The closed string sector is composed of two Kahler moduli $T_1$, $T_2$, the dilaton $S$ and some set of complex-structure moduli $U$. Note that in the toy model we have taken the modular weight of the open string fields to be zero. 

This corresponds to realising the ISS sector on a D7 brane setup where the (inverse squared) gauge coupling is given by $\im{T_1}$. A mirror IIA construction would have a D6 brane realisation. We may also realise such a model by replacing $T_1$ with the dilaton $S$ which is related to the gauge coupling of D3 branes \cite{Angelantonj:2007ts}. Either way there is an important point to raise here. Such theories may or may not be anomalous with mixed anomalies that may need to be cancelled through the GS mechanism as described in section \ref{chirmat}. We have already argued that in that case the modulus can not appear perturbatively in the superpotential. Hence a model such as (\ref{isstoy}) where the gauge coupling modulus appears perturbatively requires a realisation of the gauge sector which is anomaly free in that sense. This could arise for example from non-chiral constructions such as in \cite{Landsteiner:1998pb}. With this assumption made clear we proceed to study the vacuum structure of this model. 

We can now integrate out all the string scale moduli that appear perturbatively by replacing them with values as fixed by $W^P$. This leaves an effective low energy theory 
\ba
K &=& -2 \mathrm{ln\;}\Vol_0 + \left| \vp \right|^2 + \left| \tilde{\vp} \right|^2 + \left| \Phi \right|^2 \;, \nn \\
W &=& W_0 + h\mathrm{Tr\;} \tilde{\vp}\Phi\vp - h\mu^2\mathrm{Tr\;} \Phi  \;, \label{isstoyeff}
\ea
where $\Vol_0$ is the volume of the manifold in the vacuum and $\mu^2=\left<mAe^{iaT_1}\right>$. 
Note that both $W_0$ and $\mu$ are dynamically generated and so can naturally take small values.
It is simple to show, see \cite{Angelantonj:2007ts}, that the $W_0$ term in the superpotential and the ISS theory of the last two terms decouple as long as $\mu$ is small and $W_0 \sim \mu^2$.
In that case we find a scalar potential
\be
V = \frac{1}{\Vol_0^2}\left[ \left( \left|\partial_{\vp}W\right|^2 + \left|\partial_{\tilde{\vp}}W\right|^2 + \left|\partial_{\Phi}W\right|^2 \right) - 3\left| W_0 \right|^2 \right]+ \O\left( \mu^5 \right)\;, \label{isstoyveff}
\ee
Then it was shown in \cite{Intriligator:2006dd} that the first term in (\ref{isstoyveff}) has a metastable minimum at 
\be
\Phi_0 = 0 \;\;,\;\; \vp_0 = \tilde{\vp}_0^T = \left( \begin{array}{cc} \mu I_n \\ 0 \end{array} \right) \;,
\ee
where $I_n$ is the $n \times n$ identity matrix. In this vacuum we get
\be
V = \frac{1}{\Vol_0^2}\left[ \left( N_f - n \right)h^2\mu^4 - 3\left| W_0 \right|^2 \right] + \O\left( \mu^5 \right)\;. \label{isstoyveff2}
\ee
Then $W_0$ acts as an effective cosmological constant which can be tuned to cancel the first term. This requires $W_0 \sim \mu^2$ as assumed in deriving the effective theory. This produces a non-supersymmetric vacuum with $m_{\frac32} \sim W_0$ which requires $W_0 \sim \mu^2 \sim 10^{-13}$ for TeV supersymmetry breaking. 

A similar scenario was studied in \cite{Angelantonj:2007ts} where they considered the ISS theory to be realised on D3 branes and fixed the dilaton perturbatively using normal fluxes. We see that non-geometric fluxes allow us to also realise the gauge theory on D6 or D7 branes. 

%%%%%%%%%%%%%%%%%%%%%%%%%%%%%%%%%%%%%%%%%%%%%%%%%%%%%%%%%%%%%%%%%%%%%%%%%%%%%%%%%%%%%%%%%%%%%%%%%%%%%%%%%
\section{Summary}
\label{summ}
%%%%%%%%%%%%%%%%%%%%%%%%%%%%%%%%%%%%%%%%%%%%%%%%%%%%%%%%%%%%%%%%%%%%%%%%%%%%%%%%%%%%%%%%%%%%%%%%%%%%%%%%%%

In this paper we studied some of the issues related to the introduction of non-geometric fluxes. We studied models that are completely mirror 
symmetric and so the results apply to IIA and IIB. Using toy models based on CY data we were able to show that perturbatively exact supersymmetric 
Minkowski vacua with all the moduli stabilised can be found. 
These can arise with or without an associated R-symmetry. In both cases the vacua are not as rare as in the geometric landscape. We also showed 
that at least the complex-structure moduli could be parametrically controlled in these vacua, but were unable to show that this holds for all the 
moduli. 

We went on to show that the orientifold projection guarantees that non-perturbative effects in the same fields that appear perturbatively are 
consistent with supergravity gauging and world-volume anomaly constraints. These effects perturbed the supersymmetric Minkowski vacua to vacua 
with an exponentially small gravitino mass. This forms a concrete realisation of a dynamically low gravitino mass. The resulting vacua should generically be 
supersymmetric AdS but, given a lack of explicit knowledge of the Kahler potential and non-perturbative superpotential, no explicit solutions were presented.

The current constructions of chiral theories on intersecting branes require axions from the closed string sector. The implementation of such axions within 
the non-geometric constructions required switching off the fluxes associated with the cycle that the brane wraps. These lead to a generalised KKLT scenario 
where the moduli are split into a sector that is fixed perturbatively and a sector that is fixed non-perturbatively but with both sectors generally consisting of a mixture of complex-structure and Kahler moduli. 

We discussed ways to break supersymmetry that preserve the hierarchically small gravitino mass and in particular showed that metastable vacua were complimented by non-geometric fluxes. The fluxes allowed for realisations of the gauge sector on D6 and D7 branes as opposed to just D3 branes. 

If we do not require axions, then all the moduli can be fixed at a high string scale. This has a number advantages. 
It is particularly suited to realising gauge mediated supersymmetry breaking without a light moduli problem \cite{Diaconescu:2005pc,Kawano:2007ru,deAlwis:2007qx}. Also 
it could be important in realising inflation scenarios with a large amount of tensor modes. This follows since these 
are scenarios where, although the gravitino mass is at the TeV scale, the barrier from the minimum to the 
runaway vacuum is at the high string scale since the original vacuum was perturbative. This  
provides a solution to the problem pointed out in \cite{Kallosh:2007wm} where the vacuum is destabilised if the 
Hubble constant is too large during inflation. 

Perhaps the most relevant future directions within non-geometric compactifications are finding explicit examples of manifolds and constructing realistic D-branes sectors. 
In particular it would be nice to find models that do not require closed string axions to cancel anomalies as the destabilisation of the moduli this implies seems unnatural.

This paper studied constructions that can be described equally in IIA or IIB. 
There should also be dual constructions in the heterotic strings perhaps compactified on S-folds. On a more general level it would be nice to be able to study the low energy phenomenology of M-theory independently of which corner we work in. Perhaps this will lead to generic predictions from the full landscape.

%%%%%%%%%%%%%%%%%%%%%%%%%%%%%
\section*{Acknowledgements}
%%%%%%%%%%%%%%%%%%%%%%%%%%%%%

I thank Philip Candelas, Pablo Camara, Joseph Conlon, Daniel Cremades, Miriam Cvetic, James Gray, Andrei Micu, Gianmassimo Tasinato and Angel Uranga for useful discussions 
and patient explanations. I especially thank Pablo Camara, Joseph Conlon, Andrei Micu and Gianmassimo Tasinato for comments on a draft version of this paper. 
I am supported by a PPARC Postdoctoral Fellowship.

%%%%%%%%%%%%%%%%%%%%%%%%%%%%%%%%%%%%%%%%%%%%%%%%%%%%%%%%%%%%%%%%%%%%%%%%%%%%%%%%%%%%%%%%%%%%%%%%%%%%%%%%%
% BIBLIOGRAPHY
%%%%%%%%%%%%%%%%%%%%%%%%%%%%%%%%%%%%%%%%%%%%%%%%%%%%%%%%%%%%%%%%%%%%%%%%%%%%%%%%%%%%%%%%%%%%%%%%%%%%%%%%%%

\end{document}